\documentclass[prb,amsmath,twocolumn,showpacs,superscriptaddress]{revtex4}
\usepackage{epsfig}

\usepackage{color}
\def\beq{\begin{equation}}
\def\eeq{\end{equation}}
\def\beqn{\begin{eqnarray}}
\def\eeqn{\end{eqnarray}}
\newcommand{\be}{\begin{equation}}
\newcommand{\ee}{\end{equation}}
\newcommand{\bea}{\begin{eqnarray}}
\newcommand{\eea}{\end{eqnarray}}
\newcommand{\bphi}{{\boldsymbol {\phi}} }
\newcommand{\bsigma}{{\boldsymbol {\sigma}} }
\newcommand{\bk}{{\bf k}}
\newcommand{\br}{{\bf r}}

\newcommand{\bq}{{\mathbf q}}

\newcommand{\bQ}{{\mathbf  Q}}
\newcommand{\ud}{{\textrm{d}}}
\begin{document}
\title{Quantum order-by-disorder driven phase reconstruction in the vicinity of ferromagnetic quantum critical points}

\author{Una Karahasanovic}
\affiliation{SUPA, School of Physics and Astronomy, University of St.~Andrews, North Haugh, St. Andrews, KY16 9SS, United Kingdom}
\author{Frank Kr\"uger}
\affiliation{SUPA, School of Physics and Astronomy, University of St.~Andrews, North Haugh, St. Andrews, KY16 9SS, United Kingdom}
\author{Andrew G. Green}
\affiliation{London Centre for Nanotechnology, University College London, Gordon St., London, WC1H 0AH, United Kingdom}

\date{\today}
\begin{abstract}
The formation of new phases close to itinerant electron quantum critical points has been observed experimentally in many compounds. We present a unified analytical model that explains the emergence of new types of order around itinerant ferromagnetic quantum critical points. The central idea of our analysis is that certain Fermi-surface deformations associated with the onset of the competing order enhance the phase-space available for low-energy quantum fluctuations and so self-consistently lower the free energy. We 
demonstrate that this quantum order-by-disorder mechanism leads to instabilities towards the formation of spiral and d-wave spin nematic phases close to itinerant ferromagnetic quantum critical points in three spatial dimensions.
\end{abstract}


\maketitle
\section{Introduction}
A variety of unusual phenomena that do not fit into the framework of Fermi-liquid theory have been observed in the vicinity of quantum phase transitions. 
An interesting example is the emergence of new phases near to putative quantum critical points, observed in many experiments.
Examples include the onset of superconductivity close to the itinerant ferromagnetic quantum critical point of UGe$_2$, \cite{Saxena+00,Huxley+02} an anomalous anisotropic phase around the metamagnetic quantum critical end point of Sr$_3$Ru$_2$O$_7$, \cite{Borzi+07,Rost+09} a possible inhomogeneous magnetic state in ZrZn$_2$, \cite{Uhlarz+04} or the unusual partially ordered phase of MnSi.\cite{Pfleiderer+04} This has led to the speculation that the onset of new phases might represent a generic principle.\cite{Laughlin+01}

In his pioneering work, Hertz \cite{Hertz76} studied the paramagnet-to-ferromagnet quantum phase transition of itinerant fermions that occurs by varying the exchange coupling between electron spins. He derived an effective action for dynamical fluctuations of the bosonic order parameter. Later, Millis \cite{Millis93} used this approach to calculate temperature dependencies of the correlation length, susceptibility and specific heat. In the past decade several authors \cite{Belitz+97,Betouras+05,Rech+06,Efremov+08,Maslov+09} carried out diagrammatic calculations that extend beyond Hertz-Millis theory. They showed that the free energy contains a non-analytic dependence on the order parameter and its gradients. It was argued \cite{Belitz+97} that non-analytic terms occur due to additional soft particle-hole modes that couple to the order parameter. These non-analytic corrections can render the transitions weakly first order at low temperatures and lead to the instability of quantum critical points to the formation of new phases. 

We present an alternative, analytical approach - \emph{quantum order-by-disorder} -  which is able to predict new phases and provides an intuitive physical picture of the problem. Moreover it results in relatively simple calculations, accessible to a broad audience. It relies on the idea that certain deformations of the Fermi surface enhance the phase space available for quantum fluctuations and thus self-consistently lower the free energy. This results in new phases near to the putative quantum critical point. Note that in the familiar realizations,\cite{Mila+91,Zaanen00,Zaanen+01,Kruger+06} the quantum order-by-disorder mechanism is driven by bosonic order-parameter fluctuations. Here, the underlying fermionic statistics of the excitations and Pauli blocking of phase space is very important. Recently, it has been demonstrated\cite{Kruger+12} that the formation of the enigmatic partially ordered phase of MnSi\cite{Pfleiderer+04} with its peculiar magnetic-ordering pattern can be explained by the fermionic quantum order-by-disorder mechanism.

Our work builds upon the work of Conduit \emph{et al},\cite{Conduit+09} in which a spiral phase was predicted close to the itinerant ferromagnetic quantum critical point in three spatial dimensions. In that work, the quantum order-by-disorder approach was used with a numerical evaluation of the fluctuation corrections to the free energy in the presence of a spiral state.  We develop an analytical approach that ultimately allows us to extend the framework to include new phases such as the spin nematic, a Pomeranchuk-type instability in which the net magnetization is absent and the Fermi-surface deformations for spin-up and spin-down electrons are of opposite sign.\cite{Hirsch90,Wu+07,Chubukov+09}

This is achieved through a Ginzburg-Landau expansion in the vicinity of the finite temperature tricritical point.
We calculate closed expression for the Ginzburg-Landau coefficients of a uniform ferromagnet and evaluate them analytically in the limit $T \rightarrow 0$.
The Ginzburg-Landau coefficients of the spiral ferromagnet are related to the coefficients of uniform ferromagnet by averages of certain angular functions.

For the spin nematic we first develop an expansion of the generating function in powers of the field conjugate to the spin-nematic order parameter. Similarly, the 
coefficients in this expansion are related to uniform Ginzburg-Landau coefficients by averages of certain angular functions. A Legendre transform of the generating 
function recovers the expression for the free energy. By including small deviations from the isotropic free-electron dispersion, we are able to obtain phase diagrams 
relevant to a broader range of experimental systems. 

This paper is organized as follows: In Section II we discuss the key ideas of quantum order-by-disorder and outline its mathematical setting. In Section III we proceed to construct 
the Ginzburg-Landau expansions for the uniform ferromagnetic, spin-spiral, and spin-nematic states.
This enables us to construct a phase diagram in Section IV. Finally, in Section V we summarize our results and suggest directions for future work.  

\section{Quantum Order-by-Disorder}

The central idea of quantum order-by-disorder is that certain phases have more low-energy quantum fluctuations associated with them. This lowers their free energy and renders them stable. The effect of the lowering of the free energy already becomes evident in second order perturbation theory; the second order contribution to the free energy of the ground-state is always negative. The mechanism is similar in some ways to the entropic lowering of the free energy in classical systems.  Equivalent results for the contribution of quantum fluctuations to the free energy can be derived starting from a functional integral approach.

A well known example of quantum order-by-disorder is that of a quantum antiferromagnet. If the electron spins are oriented ferromagnetically, no virtual electron hopping is allowed 
due to the Pauli exclusion principle. On the other hand, when the spins are antiparallel, electron hopping is allowed. This hopping lowers the free energy of the system through 
second order perturbation theory. In this way the antiferromagnetic phase is stabilized due to the effect of quantum fluctuations.

\subsection{Perturbation Theory}
Let us now begin to develop this general heuristic picture into an explicit calculation. 
Our starting points is the free electron system in $d=3$ spatial dimensions interacting through Hubbard point repulsion 
\begin{equation}
\label{Hamiltonian}
\mathcal{H}=\sum_{{\bf k}, \sigma=\pm}  \left (\epsilon_{\bk}-\mu \right )\hat n_{{\bf k},\sigma}+g\int \ud^3{\bf r} \phantom{.} \hat n_{+}({\bf r})\hat n_{-}({\bf r}).
\end{equation}
Here $\epsilon_{\bk}=\frac{k^2}{2}$ is the isotropic free-electron dispersion, $\mu$ denotes the chemical potential and $\hat n_{\pm}({\bf r})$ densities of spin up/down electrons. 
Note that later on, we will include small anisotropic deformations to make the dispersion more tight-binding like. The strength of the contact interaction is given by $g$.
The mean-field free energy is given by
\begin{eqnarray}
\label{mffreeenergy}
\mathcal{F}_{\textrm{MF}}&=&-\frac{1}{\beta} \sum_{{\bf k}, \sigma} \ln{(1+e^{-\beta(\epsilon^\sigma_{{\bf k}}-\mu)})} +g \int \ud^3\br M^2(\br),
\end{eqnarray}
where $\beta$ represents the inverse temperature, $\epsilon^\sigma_{{\bf k}}$ is the mean-field dispersion in the presence of certain type of order, and ${\bf M}(\br)$ is the  magnetization vector. In this work we will not consider phases with spatial charge modulations.

The effects of fluctuations are already seen in self-consistent second order perturbation theory. The fluctuation corrections to the free energy are given by
\begin{equation}
\label{uncorrectedfreeenergy}
\tilde{\mathcal{F}}_{\textrm{fl}} = - 2 g^2 \sum'_{\bk_1\ldots\bk_4}\frac{n^+_{\bk_1}n^-_{\bk_2}(1-n^+_{\bk_3})(1-n^-_{\bk_4})}{\epsilon^+_{\bk_1}+\epsilon^-_{\bk_2}-\epsilon^+_{\bk_3}-\epsilon^-_{\bk_4}}, 
\end{equation}
where the prime over the sum indicates momentum conservation, $\bk_1+\bk_2=\bk_3+\bk_4$, and for brevity, we have written the Fermi functions as
\begin{equation}
n^\sigma_\bk:=n(\epsilon^\sigma_\bk)=(e^{\beta{(\epsilon^\sigma_{\bk}-\mu)}}+1)^{-1}.
\end{equation}

Note that the fluctuation corrections to the free energy are calculated self-consistently; the energies entering the Fermi functions are the mean-field dispersions in the presence of a given type of order. 

\begin{figure}
\includegraphics[width=0.6 \linewidth]{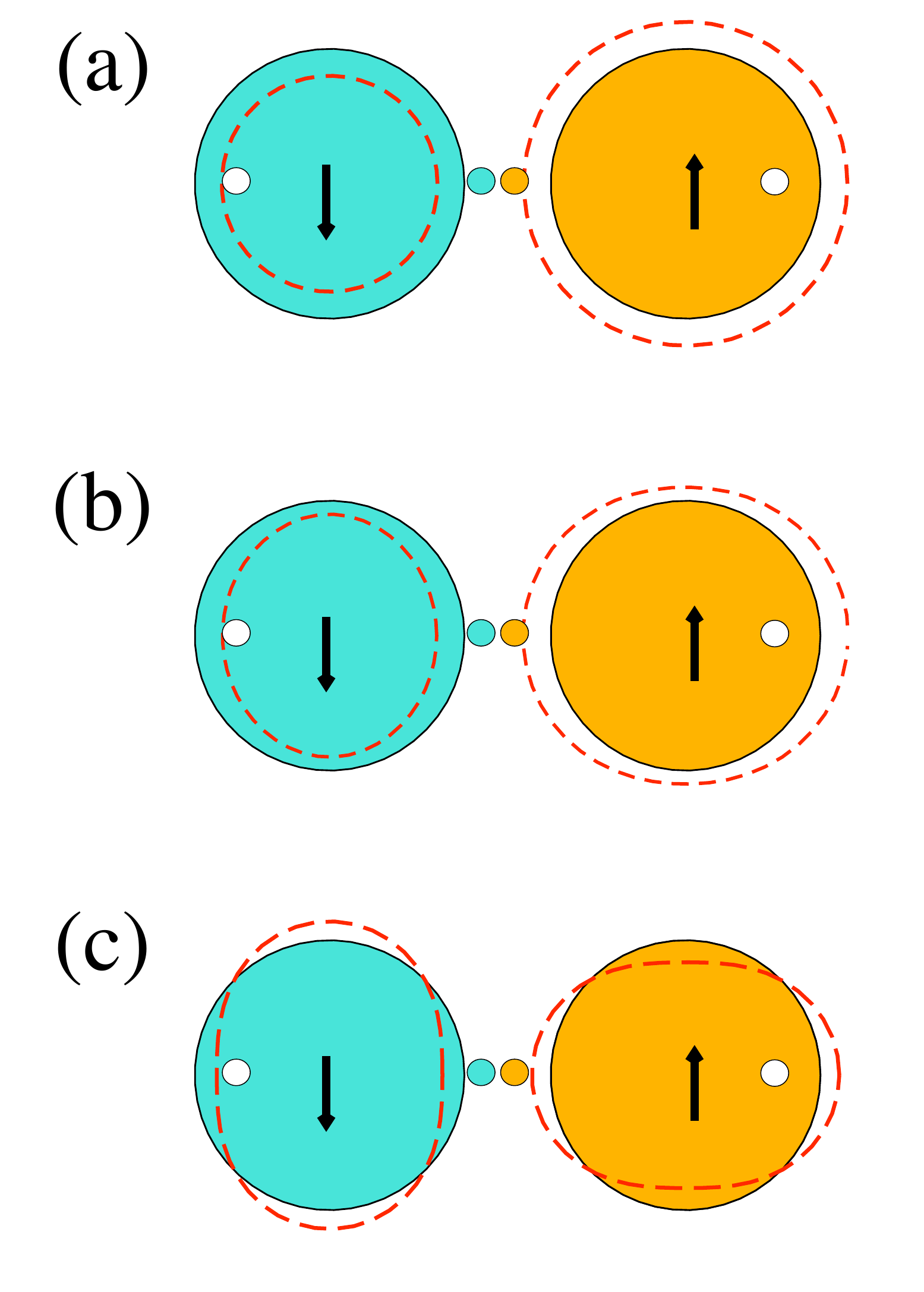}
\caption{(Color online) Distortions of the Fermi surface (dashed lines) enhance the phase-space for quantum fluctuations. (a) uniform ferromagnet (b) spiral and (c) d-wave spin nematic. Quantum fluctuations correspond to excitations of pairs of particle-hole pairs of opposite spin and equal and opposite momenta. }
\label{distortions}
\end{figure}

From Eq.(\ref{uncorrectedfreeenergy}) we see that the fluctuations correspond to excitations of virtual pairs of particle-hole pairs of opposite spin and equal and opposite momenta (spin up particle-hole pairs carry momentum $\bk_1-\bk_3$ and spin down particle-hole pairs carry momentum $\bk_2-\bk_4$). Since we need to put in energy to create the particle-hole pairs, the denominator of  (\ref{uncorrectedfreeenergy}) is always positive, which results in \emph{negative} contributions to the free energy. Certain deformations of the 
Fermi-surface enhance the phase space available for low energy, virtual particle-hole excitations and in that way self-consistently stabilize new phases. Ferromagnetic, spiral or spin-nematic Fermi-surface distortions which are shown schematically in Fig.~(\ref{distortions}) all open up extra phase-space for the low-energy 
particle-hole pairs to form. 

Careful inspection of Eq.~(\ref{uncorrectedfreeenergy}) reveals that the term 
\begin{equation}
\tilde{\mathcal{F}}_{\textrm{fl}}^\infty = - 2 g^2 \sum'_{\bk_1\ldots\bk_4}\frac{n^+_{\bk_1}n^-_{\bk_2}}{\epsilon^+_{\bk_1}+\epsilon^-_{\bk_2}-\epsilon^+_{\bk_3}-\epsilon^-_{\bk_4}}
\end{equation}
contained in $\tilde{\mathcal{F}}_{\textrm{fl}}$ gives an unphysical divergent contribution to the free energy. To avoid this, we need to take into account the renormalization of the interaction matrix element $g$.\cite{Pathria96} We perform a self-consistent perturbative expansion around a mean-field solution. Instead of using the momentum to label the eigenstates of the free-electron Hamiltonian, we use it to label first-order shifted states,
\begin{eqnarray}
\vert {\bf k}\uparrow,{\bf l}\downarrow \rangle
&=&
\vert {\bf k}\uparrow,{\bf l}\downarrow \rangle_0
\nonumber\\
& &
+
\sum_{{\bf p} \ne {\bf k}, {\bf q}\ne {\bf l}}
\frac{
_0 \langle {\bf p}\uparrow,{\bf q}\downarrow \vert
{\cal H}^{\textrm{int}}
\vert {\bf k}\uparrow,{\bf l}\downarrow \rangle_0
}
{
\epsilon^+_{\bf k}+\epsilon^-_{\bf l}-\epsilon^+_{\bf p}-\epsilon^-_{\bf q}
}
\vert {\bf p}\uparrow,{\bf q}\downarrow \rangle_0
\nonumber\\
\end{eqnarray}
where
$\vert{\bf k}\uparrow,{\bf l}\downarrow \rangle_0$ labels the two-particle free electron state,
$\vert{\bf k}\uparrow,{\bf l}\downarrow \rangle$ labels the first order corrected two-particle state, $\epsilon^\sigma_{\bf k}$ are taken self-consistently to be the mean-field electron dispersions and ${\cal H}^{\textrm{int}}$ represents the interaction Hamiltonian.
With this identification, we must also make a corresponding alteration to the interaction strength,
\begin{equation}
\label{grenormalization}
g_{\bk_1,\bk_2} \rightarrow g-2 g^2 \sum '_{\bk_3,\bk_4} \frac{ 1 }{\epsilon^+_{\bk_1}+\epsilon^-_{\bk_2}-\epsilon^+_{\bk_3}-\epsilon^-_{\bk_4}}.
\end{equation}

This renormalization corresponds to a one-loop correction to the interaction strength and leads to a regular expression for the free energy,
\begin{equation}
\label{freeenergyn}
\mathcal{F}_{\textrm{fl}} =  2 g^2 \sum'_{\bk_1\ldots\bk_4}\frac{n^+_{\bk_1}n^-_{\bk_2}(n^+_{\bk_3}+n^-_{\bk_4})}{\epsilon^+_{\bk_1}+\epsilon^-_{\bk_2}-\epsilon^+_{\bk_3}-\epsilon^-_{\bk_4}}.
\end{equation}

\subsection{Functional Integral Derivation}

We next sketch how the same result can be derived using field theoretical methods.\cite{Conduit+09,Conduit+09b}
This approach reveals immediate connection between the second order self-consistent perturbation theory, outlined above, and field theoretical calculations that explicitly show non-analytic behavior of the free energy. We start from the fermionic partition function,
\begin{eqnarray}
\mathcal{Z}&=&\int \mathcal{D}(\bar \psi, \psi)e^{-{\cal S}[\bar \psi, \psi] }, \nonumber \\
{\cal S}[\bar \psi, \psi]&=&\int^{\beta}_{0}\ud\tau \int \ud^3 {\bf r}\left [  \bar \psi \partial_{\tau}\psi +\mathcal{H}(\bar \psi,\psi)    \right ],
\end{eqnarray}
where $\psi=(\psi_+,\psi_-)^T$ and $\bar \psi=(\bar\psi_+,\bar\psi_-)$ denote Grassman fields which vary throughout space and imaginary time, and the Hamiltonian is given by Eq.~(\ref{Hamiltonian}).
After performing a Hubbard-Stratonovich decoupling of the interaction in spin ($\boldsymbol{\phi}$) and charge ($\rho$) channels we obtain
\begin{eqnarray}
{\cal Z}
&=&
\int \mathcal{D}(\bar \psi, \psi) \mathcal{D} {\boldsymbol {\phi}}  \mathcal{D} \rho 
e^{-{\cal S}[\bar \psi, \psi, \bphi,\rho]}, \nonumber
\\
{\cal S}[\bar \psi, \psi, \bphi, \rho]
&=&
\int\bar \psi( {\hat G_0^{-1}} +g(\rho-\bphi  \cdot\boldsymbol{\sigma}) ) \psi \nonumber \\
 &&+g\int
(\bphi^2-\rho^2) ,
\end{eqnarray}
where ${\hat G_0^{-1}}$ is the free-electron Green function and $\boldsymbol{\sigma}$ denotes the vector of Pauli matrices.
Integrating out the fermions, we obtain
\begin{eqnarray}
{\cal Z}
&=&
\int \mathcal{D} {\boldsymbol {\phi}}  \mathcal{D} \rho 
e^{-{\cal S}[\bphi,\rho]},
\\
{\cal S}[{\bphi}, \rho] \nonumber
&=&
- \textrm{Tr}   \ln
 \left[
 {\hat G_0^{-1}}
 +g(\rho-\boldsymbol{\sigma} \cdot {\boldsymbol \phi })\right ] 
 + g\int
(\bphi^2-\rho^2) .
\end{eqnarray}

So far, all the steps are the same as in Hertz-Millis theory. However, in that case the aim was to derive an effective action for dynamical fluctuations of the bosonic order parameter in the paramagnetic state. In contrast, we wish to derive a Ginzburg-Landau expansion in the static order parameter. In order to do this we separate $\bphi$ and $\rho$ into zero-frequency $(\rho_0,{\bf M})$ and finite-frequency parts $(\tilde \rho,\tilde \bphi)$:  $\rho=\rho_0+\tilde \rho$ and $\bphi={\bf M}+\tilde \bphi$. The action then becomes
\begin{eqnarray}
{\cal S}[{\bf \phi}, \rho]
& = & -  \textrm{Tr}  \ln
 \left[
 {\hat G_0^{-1}}+g\rho_0-g \bsigma . {\bf M} 
 + g(\tilde\rho-\tilde \bphi \cdot\bsigma) \right ] \nonumber \\
 &+& g\int ({\bf M}^2+\tilde \bphi ^2-\tilde \rho^2 ).
\end{eqnarray}
 We expand this action to quadratic order in finite-frequency fluctuations and integrate them out, yielding the following expression for the free energy:
\begin{eqnarray}
\mathcal{F}[{\bf M}]&=&g M^2-\textrm{Tr} \ln{{\hat G^{-1}}} \nonumber \\&&+\underbrace{\frac{1}{2}\textrm{Tr} \ln (1+2 g \Pi^{+-} +g^2 \Pi^{+-}\Pi^{-+})}_{\mathcal{F}_{\textrm{fl},\perp}} \nonumber \\&&+\underbrace{\frac{1}{2}\textrm{Tr} \ln (1-g^2 \Pi^{++}\Pi^{--})}_{\mathcal{F}_{\textrm{fl},\parallel}}.
\end{eqnarray}
The term $\mathcal{F}_{\textrm{fl},\perp}$ arises from transverse magnetic fluctuations, while the term $\mathcal{F}_{\textrm{fl},\parallel}$ accounts for charge-density and 
longitudinal magnetic fluctuations.
The polarization bubbles are given by
\begin{equation}
\Pi^{\sigma \sigma '}({\bf q},\omega)=\frac{1}{\beta}\sum_{\bk,\omega'}\hat{G}_\sigma (\bk,\omega')\hat{G}_{\sigma'}({\bf k}-{\bf q},\omega'-\omega),
\end{equation}
where $\hat G_{\sigma}^{-1}=\hat G_0^{-1}+g\rho_0-\sigma g M $.
The polarization bubbles $\Pi^{\sigma \sigma '}$ explicitly depend on the interaction strength $g$. 

Next, we perform a rather strange expansion in $g$. We expand the $\textrm{Tr}\ln$-terms to second order in $g$ by only expanding in powers of $g$ that stand in front of the polarization bubbles, while keeping the full $g$ dependence of the polarization bubbles as it is. This looks like a second order expansion in $g$, but self-consistency actually implies resummation of certain classes of contributions to infinite order. This expansion captures the relevant physics, as we will see later on. After performing the summations over Matsubara frequencies, we arrive at the expression (\ref{uncorrectedfreeenergy}). Further, we need to renormalize $g$, to cancel the ultraviolet divergence. Doing so according to Eq.~(\ref{grenormalization}) recovers expression Eq.~(\ref{freeenergyn}) for the free energy. \\

In summary, quantum order-by-disorder provides a physical picture for the formation of new phases due to quantum fluctuations. As we will see later, certain deformations of the Fermi surface enhance the phase space available for quantum fluctuations and in that way lower the free energy. This is already apparent in second order perturbation theory 
and can also be derived from a functional integral approach. Next, we want to expand the free energy in powers of the order parameter (which enters through the mean-field dispersion). This will enable us to construct the Ginzburg-Landau expansion and to analyze the phase diagram.

\section{Ginzburg-Landau Expansion}

We wish to determine the phase diagram of the near critical itinerant ferromagnet allowing for the generation of new phases near the quantum critical point. In order to obtain the phase diagram, we develop a Ginzburg-Landau expansion of the free energy in powers of the order parameters for the various  types of phases that might form. The expansion is valid in the vicinity of the tricritical point (explained later), where the value of the order parameters is sufficiently small.

It turns out that our task is simplified considerably by relationships between the expansion coefficients for the different types of order and those for the uniform ferromagnet. We begin therefore with an explicit evaluation of the coefficients for the uniform ferromagnet. 

Next, we allow for spatial modulations of the ferromagnetic order; in particular we consider a spiral modulation of the magnetization. We use the fact that the free energy can be expressed (to all orders) as a functional of the mean-field electron dispersion in the presence of the spiral order. We show how the coefficients of the expansion in the spiral 
ordering wave vector ${\bf Q}$  are related (by angular averages of certain functions) to those of the uniform ferromagnet. 

For other order parameters, not driven in mean-field, we introduce a field conjugate to the order parameter and construct an expansion of the generating function in terms of the conjugate field. We are able to relate the coefficients of the generating function to the Ginzburg-Landau coefficients of a uniform ferromagnet by performing some simple angular integrals. We use a Legendre transformation of the generating function to recover the Ginzburg-Landau expansion. Quantum fluctuations generate an interaction in the new channel. A similar mechanism is well known in spin-fluctuation theory where superconductivity is mediated through spin-fluctuations.\cite{Anderson+73,Fay+80} In what follows, we will concentrate on the case of a $d$-wave spin nematic.

Finally, we allow for a more generic energy dispersion by considering small anisotropic deviations from the isotropic free-electron dispersion. 
We calculate the corrections to the coefficients of the Ginzburg-Landau expansion due to the anisotropic distortion. The coefficients of this expansion are proportional 
to parts of the Ginzburg-Landau coefficients of the uniform ferromagnet in the presence of an isotropic dispersion. The proportionality factors are 
determined by angular averages of functions that encode the specific form of the deviation from the isotropic free-electron dispersion.

\subsection{Uniform Ferromagnet}
The dispersion of the uniform ferromagnet  is given by $\epsilon_\sigma(\bk)=\frac{\bk^2}{2}-\sigma g M$. We Taylor expand the free energy in powers of $M$,
 \begin{equation}
 \label{hfmGL}
\mathcal{F}[M]= \alpha M^2 +\beta M^4 +\gamma M^6+...,
\end{equation}
where the Ginzburg-Landau coefficients are functions of interaction strength and temperature, $\alpha=\alpha (g,T)$, and similarly for $\beta$ and $\gamma$.

\subsubsection{Mean-Field Coefficients}
The expansion of the mean-field free energy Eq. (\ref{mffreeenergy}) in powers of $M$ leads to the following coefficients:
\begin{eqnarray}
\label{mfhomcoeff}
\alpha_{MF}&=& g + g^2 \sum_{\bk}n^{(1)}(\epsilon_{\bk}), \nonumber \\
\beta_{MF}&=&\frac{2}{4 !} g^4 \sum_{\bk}n^{(3)}(\epsilon_{\bk}), \nonumber \\
\gamma_{MF}&=&\frac{2}{6 !} g^6 \sum_{\bk}n^{(5)}(\epsilon_{\bk}).
\end{eqnarray}
The remaining integrals over derivatives of Fermi functions are straightforward to compute for the present $k^2$ dispersion.

\subsubsection{Fluctuation Contributions to Coefficients}
Here we outline the main steps in the calculation of the fluctuation contribution to the Ginzburg-Landau coefficients. The detailed calculation is given in the appendix. We start by writing the fluctuation corrections to the free energy in terms of modified particle-hole densities of states, Eq. (\ref{freeenergymodifieddos}). This is possible because the fluctuations 
correspond to excitations of virtual pairs of particle-hole pairs. The particle-hole densities of states can be calculated analytically as functions of temperature and magnetization 
[see Appendix A]. We can write down the expression for the fluctuation contributions to the Ginzburg-Landau coefficients, $\alpha_
\textrm{fl}$ and $\beta_\textrm{fl}$ in terms of integrals over the modified particle-hole densities of states and their derivatives with respect to the magnetization [see Appendix B]. 
The fluctuation corrections $\alpha_\textrm{fl}$ and $\beta_\textrm{fl}$ are computed analytically at low temperatures and numerically over the full temperature range.
In this work, the phase diagrams are calculated using only the mean-field contribution to the 6th order coefficient $\gamma$, since it is a higher order term in the Ginzburg-Landau expansion and since the fluctuation corrections to $\gamma$ are extremely difficult to compute. At low temperatures, the fluctuation contributions to the coefficients are given by
\begin{eqnarray}
\label{homcoeff}
\alpha_\textrm{fl} &\simeq &-\lambda (1+2\ln2)g^4, \nonumber \\
\beta_\textrm{fl} &\simeq & \lambda \left (1+\ln{\frac{T}{\mu}}\right )g^6,
\end{eqnarray}
with  $\lambda=[16\sqrt{2}]/[3 (2\pi)^6]$. Note that here and in the following, $g$ is given in dimensionless units. 
The $\ln (T/\mu)$ dependence of $\beta$ is a remnant of the $M^4\ln[M^2+\left(T/\mu \right )^2]$ term in the free energy of Belitz \emph{et al}.\cite{Belitz+97}

\subsection{Spiral}
Next, we calculate the coefficients of the Ginzburg-Landau expansion allowing for a spatial modulation of the magnetic order. We restrict our consideration to a single planar spiral.  We exploit the fact that the free energy is a functional of the mean-field dispersion in the presence of spiral magnetic order.

\subsubsection{Mean-Field Dispersion in the Presence of Spiral Magnetic Order}
First, we determine the mean-field dispersion in the presence of spiral magnetic order. Let the directrix of the spiral wave-vector point along the $z$ direction.  For a planar spiral, the magnetization vector is then restricted to lie in the $xy$ plane, ${\bf M}(\br)=M(\cos{\bQ\cdot\br},\sin{\bQ\cdot\br},0)$. Note that the Hamiltonian (\ref{Hamiltonian}) does not favor a particular direction of the spiral.  The mean-field Hamiltonian is then given by
\begin{eqnarray}
\mathcal{H}
=
\sum_{\bk}
\tilde \psi_{\bf k}^\dagger
\left( \begin{array}{cc}
 \epsilon_{\bk+\frac{\bQ}{2}}& gM \\
gM & \epsilon_{\bk-\frac{\bQ}{2}}  
\end{array} \right)
\tilde \psi_{\bf k}
+gM^2,
\end{eqnarray}
where
\begin{eqnarray}
\tilde \psi_{{\bf k}}^\dagger
&=&
\left(
\psi_{{\bf k}+{\bf Q}/2, \uparrow}^\dagger,
\;
\psi_{{\bf k}-{\bf Q}/2, \downarrow}^\dagger
\right).
\end{eqnarray}
Diagonalization of this Hamiltonian leads to the mean-field dispersion 
\begin{eqnarray}
\label{spiraldispersiongeneral}
\epsilon^\sigma_\bk & = & \frac{\epsilon_{\bk-\frac{\bQ}{2}}+\epsilon_{\bk+\frac{\bQ}{2}}}{2}\nonumber\\
& & -\sigma\sqrt{\left(\frac{\epsilon_{\bk-\frac{\bQ}{2}}-\epsilon_{\bk+\frac{\bQ}{2}}}{2}\right )^2+(gM)^2}.
\end{eqnarray}
For the case of a quadratic dispersion, this reduces to 
\begin{equation}
\label{spiraldispersion}
\epsilon^\sigma_\bk=\frac{k^2}{2}-\sigma\sqrt{(\bk\cdot\bQ)^2+(gM)^2}.
\end{equation}
We see that the spiral wave-vector enters the energy dispersion as an angle-dependent magnetization.
Note that we have absorbed a $Q^2$ term into the chemical potential.

\subsubsection{Mean-Field Ginzburg-Landau Coefficients}
We will now make use of this mean-field electron dispersion in the presence of a spiral in order to determine the Ginzburg-Landau expansion coefficients.
We Taylor expand the free energy of a spiral in powers on magnetization $M$ and wave-vector $\bQ$, keeping the terms up to $6th$ order,
\begin{eqnarray}
\label{spiralfreeprelim}
\mathcal{F}[M,Q] & = &  
\left(\alpha+\beta_{1} Q^2+\gamma_{1} Q^4\right) M^2 \nonumber\\
& & +\left(\beta+\gamma_{2} Q^2 \right) M^4 +\gamma M^6,
\end{eqnarray}
where the coefficients $\beta_{1},\gamma_{1},\gamma_{2}$ remain to be determined.  A useful simplification at this stage is to rescale the spiral wavevector according to 
$Q \rightarrow \frac{k_F}{g}Q$, so that it has the same dimensions as $M$. In this way, $\beta$ and $\beta_1$, and $\gamma$, $\gamma_1$ and $\gamma_2$ have the same dimensions. Let us first consider $\beta_1$. The mean-field contribution is given by 
$$
\beta_{1,\textrm{MF}}
=
2 \frac{2}{4 !} g^4 \int_{\bk}
\left(\frac{\bk \cdot \bQ}{k_F Q }\right)^2 n^{(3)}_{k}.
$$
 Since $T \ll \mu$, derivatives of Fermi functions are strongly peaked around the Fermi energy and we can set $\vert {\bf k} \vert =k_F$  in the scalar product which leads to a simple angular weight,  
 $$
 \frac{\bk \cdot \bQ}{k_F Q}
 \approx \cos{\theta},
 $$
  where $\theta$ is the angle between the vectors $\bk$ and $\bQ$. After carrying out the angular integral we obtain $\beta_{1,\textrm{MF}} \approx \frac{2}{3}\beta_\textrm{MF}$. Similarly, we obtain the proportionalities $\gamma_{1,\textrm{MF}}\approx \frac{3}{5}\gamma_\textrm{MF}$ and $\gamma_{2,\textrm{MF}}\approx \gamma_\textrm{MF}$. 
  
\subsubsection{Fluctuation Corrected Ginzburg-Landau Coefficients}
Now, we proceed to analyze the fluctuation corrections to the expansion coefficients in Eq.~(\ref{spiralfreeprelim}). As in the evaluation of the mean-field coefficients it turns out that the fluctuation contributions to the expansion coefficients in $Q$ are related to those of the uniform ferromagnet by angular factors which are identical to those found in the mean-field case.

The fluctuation corrections to the free energy are given by an integral over momenta $\bk_1,\ldots,\bk_4$ of a kernel that explicitly depends on each of the momenta through the mean-field dispersion (\ref{spiraldispersion}). The fluctuation contributions to the Ginzburg-Landau coefficients are obtained by differentiating Eq.~(\ref{freeenergyn}) with respect to $M$ and $Q$. First we differentiate the kernel with respect to dispersion and then the dispersion with respect to $M$ and $Q$.  
For example, the fluctuation contribution to the $M^2 Q^2$ coefficient is given by
$$\beta_{1,\textrm{fl}}
=\left.\frac{\partial^2 \mathcal{F}_{\textrm{fl}}}{\partial M^2 \partial Q^2}\right\vert_{Q=0,M=0}.
$$

We use two important facts in order to calculate this, i. that the free energy is a functional of dispersion and ii. that the spiral wave vector enters the mean-field dispersion 
Eq.~(\ref{spiraldispersion}) like an angle dependent magnetization.

The dispersion enters for each of the momenta $\bk_i$  in the momentum sum in Eq.~(\ref{freeenergyn}). Differentiating with respect to $Q^2$, therefore, will bring down factors of 
$\left(\bk_i \cdot \bQ/(k_F Q)\right)^2$,
 each of which will contribute with an angular factor as in the mean-field case. This leads to the proportionality 
$\beta_{1,\textrm{fl}}\approx \frac{2}{3} \beta_\textrm{fl}$. Combining this with the identical result for the mean-field contribution we obtain $\beta_1\approx \frac{2}{3} \beta$.
When the proportionality between all of the coefficients is taken into account the free energy (\ref{spiralfreeprelim}) becomes
\begin{eqnarray}
\label{spiralfree}
\mathcal{F}[M,Q] & = &  (\alpha +\frac{2}{3}\beta  Q^2+\frac{3}{5}\gamma Q^4) M^2\nonumber\\
& &  +(\beta +\gamma Q^2) M^4 +\gamma M^6.
\end{eqnarray}

\subsection{Other Instabilities}
Other order parameters, not driven in mean field,  are slightly more difficult to analyze. This is because we cannot decouple the interaction in those channels.
Instead, we introduce a field conjugate to the order parameter and calculate the generating function. We show how the coefficients of the generating function are 
related to the corresponding Ginzburg-Landau coefficients of the uniform ferromagnet. A Legendre transform of the generating function recovers the free energy. 
We outline this procedure for the case of a spin nematic. 

We introduce a d-wave spin-nematic order parameter 
\begin{eqnarray}
N&=&\sum_{\bk, \sigma} \sigma d_{\bk}n_{\bk,\sigma}, \nonumber \\
d_{{\bf k}}&=&\frac{1}{k_F^2}(k_x^2-k_y^2)\approx \sin^2\theta \cos(2 \phi),
\end{eqnarray}
where $d_{{\bf k}}$ is the d-wave distortion. 
The spin-nematic order parameter looks like a magnetization order parameter weighted by an angular factor. It 
corresponds to Fermi-surface distortions which have opposite signs for spin-up and spin-down electrons. The net magnetization however, vanishes 
since the volumes of the distorted spin-up and spin-down Fermi surfaces are the same (see Fig.~\ref{distortions}(c)). As we will see later, it is straightforward to generalize our final results to spin-nematic states with different symmetries. 

Turning on a fictitious field $h_{N}$  conjugate to the nematic order parameter, the dispersion is modified to $\epsilon^{\sigma}_{\bk}=\epsilon_{\bk}-\sigma h_{N} \sin^2{\theta} 
\cos(2 \phi)$. Next we expand the generating function $\varphi$ in powers of $h_{N}$,
\begin{equation}
\label{Z}
\varphi[h_N]=\alpha^\varphi h_{N}^2+\beta^\varphi h_{N}^4+\gamma^\varphi h_{N}^6,
\end{equation}
where the superscript $\varphi$ is used to distinguish coefficients of the generating function from those of the Ginzburg-Landau expansion. 

\subsubsection{Mean-field contributions to the coefficients of the generating function }
The mean-field coefficients of the generating function are given by
\begin{eqnarray}
\alpha^\varphi_{\textrm{MF}}&=&  g^2\langle d_{{\bf k}}^2 \rangle  \sum_{\bk} n^{(1)}(\epsilon_{\bk})=\langle d_{{\bf k}}^2\rangle (\alpha_{MF}-g) \nonumber \\
\beta^\varphi_{\textrm{MF}}&=&\frac{2}{4 !} g^4 \langle d_{{\bf k}}^4 \rangle \sum_{\bk} n^{(3)}(\epsilon_{\bk})=\langle d_{{\bf k}}^4\rangle \beta_{MF} \nonumber \\
\gamma^\varphi_{\textrm{MF}}&=&\frac{2}{6 !} g^6 \langle d_{{\bf k}}^6 \rangle \sum_{\bk} n^{(5)}(\epsilon_{\bk})=\langle d_{{\bf k}}^6 \rangle \gamma ,
\end{eqnarray}
where $\langle \ldots \rangle=\frac{1}{4\pi}\int_0^\pi\ud\theta \sin\theta\int_0^{2\pi}\ud\phi\ldots$ denotes the angular average. The coefficients are proportional to the corresponding uniform Ginzburg-Landau coefficients; the constants of proportionality are angular averages of powers of the nematic distortion. The resulting integrals are easy to calculate, yielding $\langle d_{{\bf k}}^2 \rangle=\frac{4}{15}$, $\langle d_{{\bf k}}^4 \rangle=\frac{16}{105}$, and $\langle d_{{\bf k}}^6 \rangle=\frac{320}{3003}$.  
Note that the term linear in $g$ in the quadratic mean-field coefficient is absent since the interaction is local in position space while the nematic distortion is local in momentum 
space, leading to a vanishing weight in the nematic channel.

\subsubsection{Fluctuation Correction to Coefficients of the Generating Function}
As we found in the case of the spiral, the fluctuation corrections to the coefficients in the nematic expansion are related to those of the uniform ferromagnet by the same angular averages as the mean-field coefficients. For example, let us consider the fluctuation contribution to the $N^4$ coefficient. When differentiating (\ref{freeenergyn}) with respect to 
$h_N$ four times, this brings down terms like $\langle d_{\bk_1} d_{\bk_2} d_{\bk_3} d_{\bk_4} \rangle$, where $\vert \bk_i \vert \approx k_F$ since derivatives of Fermi functions are peaked around the Fermi energy. Angular averages of this type are potentially more complicated as the directions of different $\bk$'s are not independent. However, the fact that the dominant contribution comes from the particle-hole pairs with momenta $\vert \bk_1-\bk_3\vert =\vert \bk_2-\bk_4\vert \approx 2 k_F$ leads to a tremendous simplification. Within this approximation, $\bk_1,\bk_2,\bk_3$ and $\bk_4$ are either parallel or antiparallel to one another, rendering 
$\langle d_{\bk_1} d_{\bk_2} d_{\bk_3} d_{\bk_4} \rangle \approx \langle d_\bk^4 \rangle$. Similar arguments hold for other types of terms that appear in the expansion. Thus, to leading order, the same proportionality holds as for the mean-field coefficients and consequently, the generating function is given by
\begin{eqnarray}
\label{nematicfreeh}
\varphi[h_N] & = & \langle d_{{\bf k}}^2 \rangle (\alpha-g) h_{N}^2  +\langle d_{{\bf k}}^4 \rangle\beta h_{N}^4 \nonumber\\
& &  +\langle d_{{\bf k}}^6 \rangle\gamma h_{N}^6.
\end{eqnarray}

\subsubsection{Ginzburg-Landau expansion of Spin Nematic}
In order to obtain the Ginzburg-Landau expansion of the free energy of the d-wave spin nematic, we perform the Legendre transform to leading order,
\begin{eqnarray}
\mathcal{F}[N]
&=&
\varphi[h_N[N]]+h_N N
\nonumber \\
\frac{\partial \varphi}{\partial h_N}
&=&
-N.
\end{eqnarray}
Carrying out this transformation, we obtain the free-energy expansion in powers of the nematic order parameter $N$,
\begin{equation}
\label{nematicfree}
\mathcal{F}[N] =- \langle d_{{\bf k}}^2 \rangle(\alpha-g) N^2  +\langle d_{{\bf k}}^4 \rangle\beta N^4  +\langle d_{{\bf k}}^6 \rangle \gamma N^6.
\end{equation}
We point out that the derivation of the free-energy functional for the spin nematic is general and not constrained to a particular symmetry of the distortion. 
To obtain the free energy of a p-wave spin nematic,\cite{Hirsch90,Wu+07,Chubukov+09} we simply replace the d-wave distortion $d_{{\bf k}}$ by the 
p-wave angular weight $p_{{\bf k}}\approx \cos\theta$. This leads to slightly different angular averages, $\langle p_{{\bf k}}^2 \rangle=\frac{1}{3}$, 
$\langle p_{{\bf k}}^4 \rangle=\frac{1}{5}$, and $\langle p_{{\bf k}}^6 \rangle=\frac{1}{7}$ and hence to slightly different coefficients in the Ginzburg-Landau expansion.

\subsection{Deviation from Free-Electron Dispersion}
Our approach also enables us to analyze the modifications to the phase diagram in the presence of a dispersion that slightly deviates from the free-electron dispersion,
$\epsilon_{\bk}=\frac{k^2}{2}+\delta \epsilon_{\bk}$. We plug this expression into the general mean-field dispersion in the presence of a spiral, Eq.~(\ref{spiraldispersiongeneral}), to obtain
\begin{equation}
\label{spiraldispersionanisotropy}
\epsilon^\sigma_{\bk}\approx\frac{k^2}{2}+\delta \epsilon_{\bk}-\sigma \sqrt{\left( \bQ\cdot(\bk+\frac{1}{2}\nabla \delta \epsilon_{\bk})\right)^2+(gM)^2}.
\end{equation}
In order to calculate the corrections to the Ginzburg-Landau coefficients, we first differentiate the free energy with respect to the dispersion and then the dispersion with respect 
to $M$ and $Q$. Finally, we expand the resulting Ginzburg-Landau coefficients in powers of $\delta \epsilon_{\bk}$, assumed small. The free energy is now given by the sum of $\mathcal{F}_{M,Q}$ (\ref{spiralfree}) and the additional contribution  
\begin{eqnarray}
\label{freehommods}
\delta \mathcal{F}[M,\bQ]
&=&
 \left( 6 \beta_{\textrm{MF}}+\frac{g^2}{2}\frac{\partial^2 \alpha_{\textrm{fl}} }{\partial^2 \mu^2} \right ) \frac{\langle \delta \epsilon_{\bk} ^2\rangle }{g^2}M^2
 \nonumber \\
&+&
15 \gamma \frac{\langle \delta \epsilon_{\bk} ^2\rangle}{g^2}M^4+ 30 \gamma \frac{\langle (\frac{\bk\cdot\bQ}{k_F})^2 \delta \epsilon_{\bk}^2\rangle}{g^2} M^2 
\nonumber \\
&+&
\frac{\beta_{\textrm{MF}}}{2} \bigg\langle \left(\frac{\bQ\cdot\nabla \delta \epsilon_{\bk}}{k_F}\right)^2 \bigg\rangle M^2 , 
\end{eqnarray}
where  $\langle ... \rangle$ denotes an angular average and we have assumed that the deviation $\delta \epsilon_{\bk}$ from the free-electron dispersion is such that the odd-power angular averages give zero.
Mixing between coefficients at different total order in $M$ and $Q$ now occurs,
 since the angular distortion  enters in both spin-symmetric and spin-antisymmetric ways, as opposed to the spin-antisymmetric Fermi-surface distortion of spiral or spin-nematic states in the isotropic case. 
 
 Similarly, the free energy of the spin-nematic state is now given by the sum of $\mathcal{F}[N]$ (\ref{nematicfree}) and an additional term given by
\begin{eqnarray}
\label{freehommodn}
\delta \mathcal{F}[N] =& -& \left ( 6 \beta_{\textrm{MF}}+\frac{g^2}{2}\frac{\partial^2 \alpha_{\textrm{fl}} }{\partial^2 \mu^2} \right )\frac{ \langle \delta \epsilon_{\bk} ^2 d_{\bk}^2 \rangle}{g^2} N ^2\nonumber \\
&+&
15 \gamma \frac{\langle \delta \epsilon_{\bk} ^2d_{\bk}^4 \rangle}{g^2} N^4. 
\end{eqnarray}

\section{Phase diagram}
In the previous section we have shown that the free energy is a functional of the mean-field dispersion.  We have used this fact to develop the Ginzburg-Landau expansions for the uniform ferromagnet and the spiral ferromagnet. The derivation of the free-energy of the spin nematic is more complicated since the point interaction has no weight in the 
spin-nematic channel and a mean-field decoupling is not possible in that case.  Instead, we added a field conjugate to the nematic order parameter, calculated the generating function and then performed a Legendre transform to obtain the Ginzburg-Landau expansion in terms of the spin-nematic order parameter. Finally, we considered small deviations from the 
isotropic free-electron dispersion and derived the resulting corrections to the free-energy functionals.

We have shown how a very useful simplification occurs in the low temperature regime; all spiral and spin-nematic coefficients can be related to those of a uniform 
ferromagnet, which we have calculated analytically at low temperatures.

We now use the Ginzburg-Landau functions developed above to construct the phase diagram as a function of temperature, $T$, and renormalized interaction 
strength, $g$. We minimize the Ginzburg-Landau free energy with respect to the order parameter(s) and compare the free energies of different phases. 
We show how quantum fluctuations stabilize the spiral and spin-nematic phases, neither of which are favored in mean-field theory. 
The effect of a small anisotropic correction to the free-electron dispersion on the topology of the phase diagram is also investigated. 

\subsection{Mean-Field Phase Diagram} 
\begin{figure}[ht]
\includegraphics[width=0.95\linewidth]{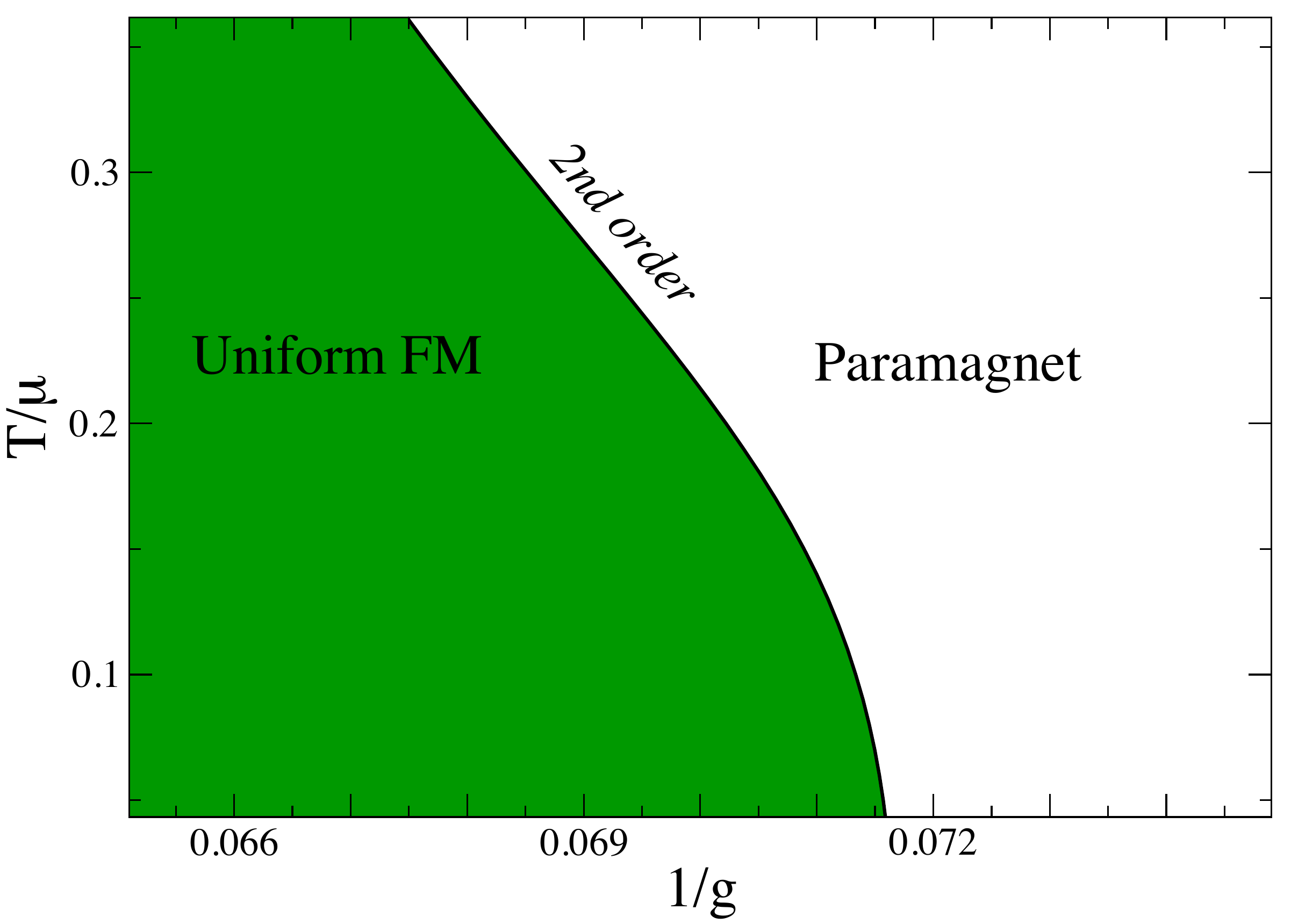}
\caption{(Color online) Phase diagram of the uniform ferromagnet in mean-field theory. The transition between the uniform ferromagnet and the paramagnet is always second order.}
\label{mfdiagram}
\end{figure}

First, we consider the mean-field phase diagram. From Eq.~(\ref{mfhomcoeff}) we see that in the low-temperature regime, the quartic coefficient, $\beta_{\textrm{MF}}$, is positive. The second order phase transition between paramagnetic and uniform ferromagnetic state happens when the quadratic coefficient, $\alpha_{\textrm{MF}}$, changes sign. 
The mean-field phase diagram is shown in  Fig.~\ref{mfdiagram}. Neither spiral, nor spin-nematic states are favored in mean field.

\subsection{Fluctuation-Corrected Phase Diagram}
Next, we include quantum fluctuations in our analysis and allow for the generation of new phases that were not present in the mean-field theory.

\subsubsection{Uniform Ferromagnet}
Before investigating how fluctuations may favor the formation of new phases, we first investigate their effect on the uniform ferromagnet. From Eq.~(\ref{homcoeff}) we see that the fluctuations provide a negative contribution to the Ginzburg-Landau coefficients. Ferromagnetism is thus favored for lower values of the interaction strength $g$ than in the mean-field theory. This becomes evident if we consider the line of second order transitions $\alpha=0$.

\begin{figure}[ht]
\includegraphics[width=0.95\linewidth]{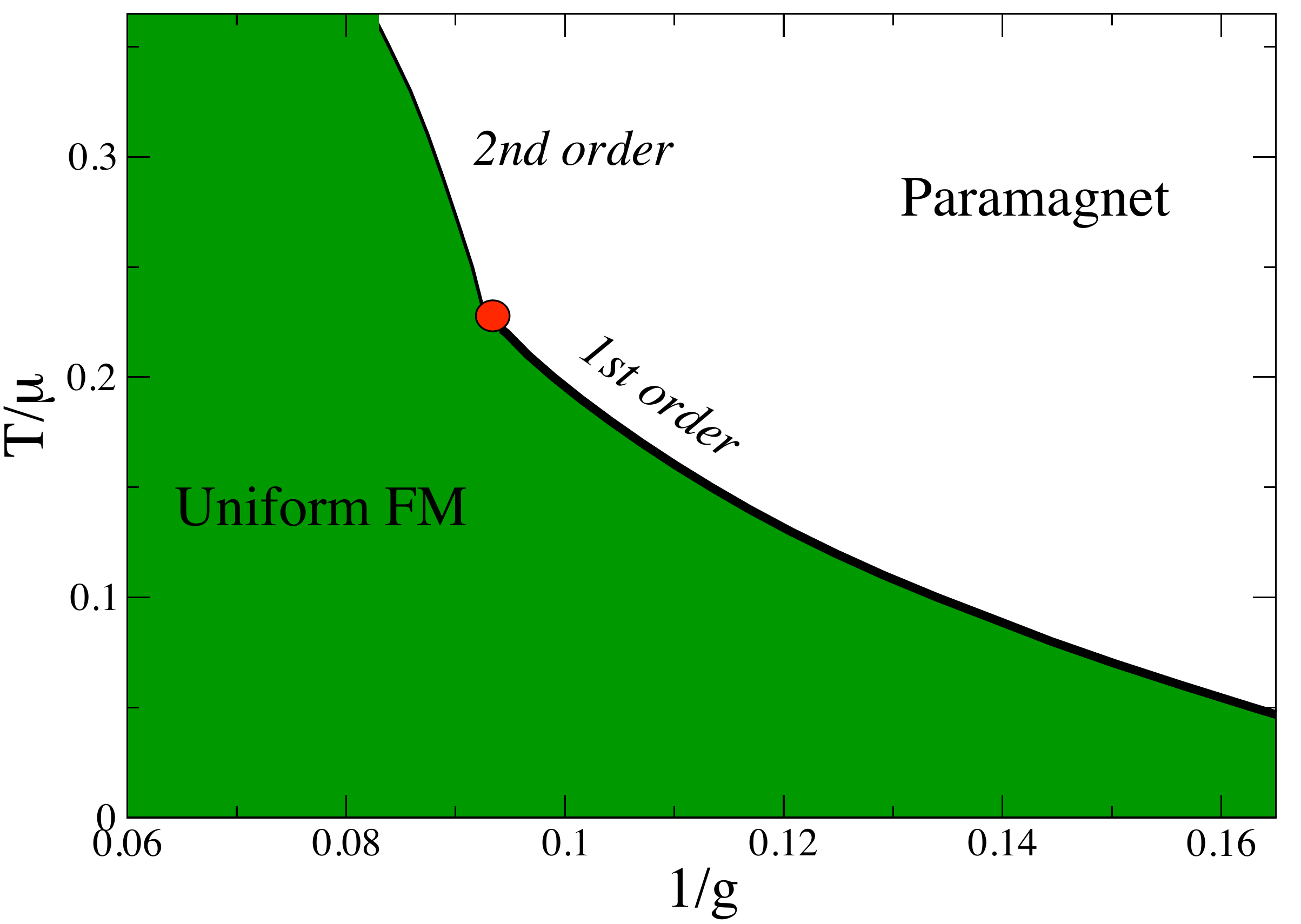}
\caption{(Color online) Phase diagram of the uniform ferromagnet, including quantum fluctuations. Below the tricritical point (shown as circle), quantum fluctuations drive 
the transition first order.}
\label{homogdiagram}
\end{figure}

In the presence of fluctuations, the quartic coefficient $\beta$ {\it inevitably} becomes negative for low enough temperatures due to the $\sim\ln(T/\mu)$ divergence. For $\beta<0$ the  paramagnet-to-ferromagnet transition becomes first order. The line of first order transitions is given by $\beta^2=4 \alpha \gamma$ (the condition for degeneracy of the minima of the Ginzburg-Landau function). The location of the tricritical point, at the intercept of the first-order and second-order lines, is found at $T^*=0.24 \mu$. This is in 
good agreement with previous numerical calculations.\cite{Conduit+09} The occurrence of first-order transitions at low temperatures has been observed in numerous experiments.\cite{Pfleiderer+97,Pfleiderer+01,Huxley+00,Yu+04,Otero+09,Uhlarz+04}

\subsubsection{Fluctuation-Driven Spiral}
\begin{figure}[ht]
\includegraphics[width=0.95\linewidth]{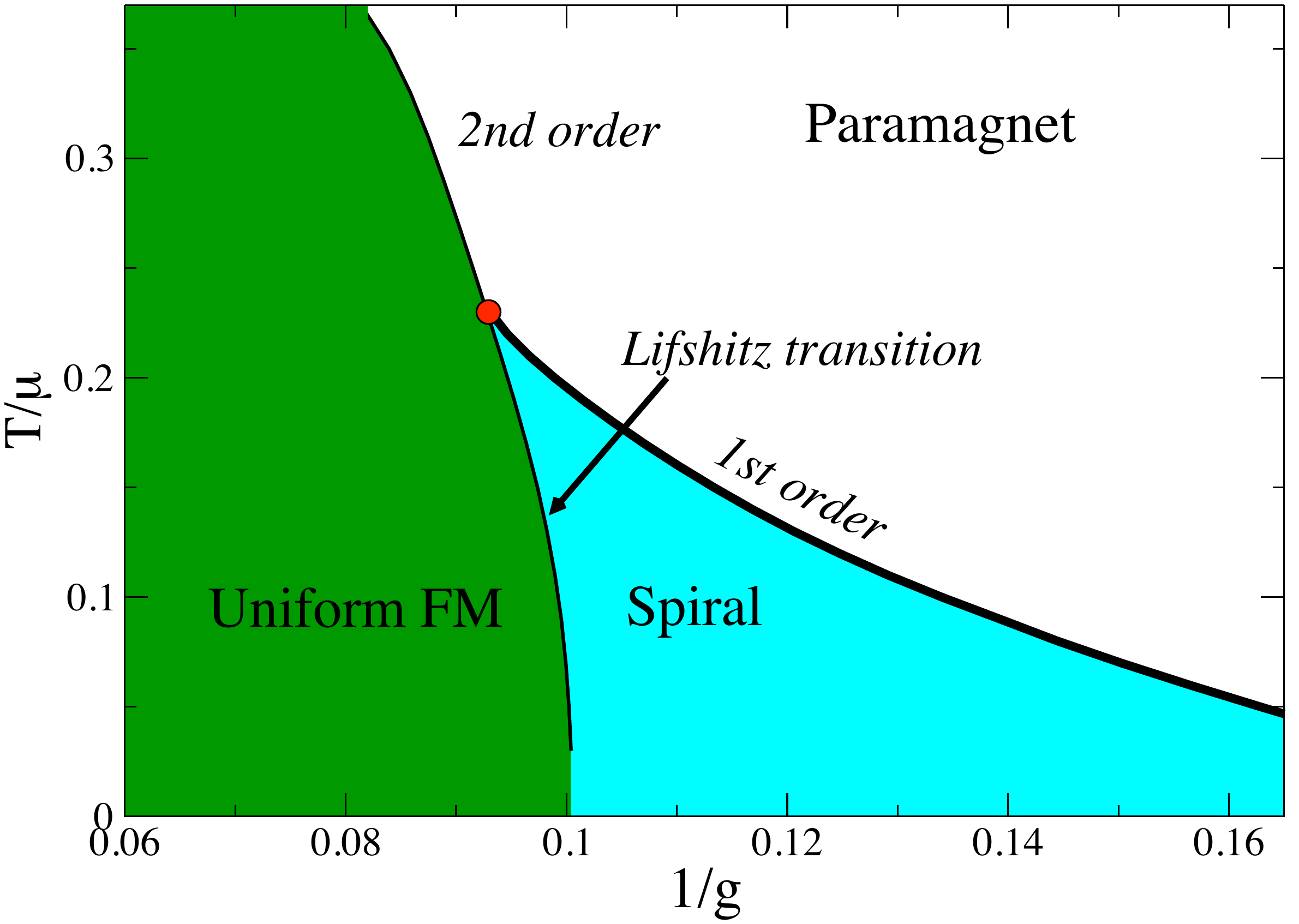}
\caption{(Color online) Phase diagram of the itinerant ferromagnet allowing for the possibility of spatially modulated ferromagnetism. Quantum fluctuations drive the 
formation of a spiral phase which sets in below the tricritical point and pre-empts the first-order transition between the uniform ferromagnet and the paramagnet. }
\label{spiraldiagram}
\end{figure}

From Eq.~(\ref{spiralfreeprelim}) we see that the $Q^2M^2$ term favors non-zero Q for $\beta_1<0$.
The particular relationship between coefficients that is found in the free-electron case implies that this occurs when $\beta < 0$, {\it i.e.} the spiral first forms at the tricritical point where the transition into a uniform magnet would have become first order. The phase diagram showing the instability towards the formation of a magnetic spiral is shown in Fig.~\ref{spiraldiagram}. We now derive it from the Ginzburg-Landau functional.
 
In the case of the free-electron dispersion we have shown that $\beta_1=2 \beta /3$. This results in the formation of a spiral state below the tricritical point. Minimizing the free energy Eq.~(\ref{spiralfree}) with respect to $Q$, we obtain the optimal wave vector
\begin{equation}
\label{QM}
\bar Q^2=-\frac{5}{6 \gamma}\left (\frac{2}{3} \beta +M^2 \gamma \right).
\end{equation}
After substituting this value of $Q$ back in (\ref{spiralfree}), we obtain the free energy at the optimal wavevector as a functional of $M$,
\begin{eqnarray}
\label{phiMspiral}
\mathcal{F}_{\bar Q}[M]
&=&
 \alpha_{\bar Q} M^2+\beta_{\bar Q} M^4 + \gamma_{\bar Q} M^6, 
 \\
 \alpha_{\bar Q} 
&=& 
\left (\alpha-\frac{5}{27}   \frac{\beta^2}{\gamma}\right ),
\nonumber \\
\beta_{\bar Q} 
&=&
\frac{4}{9}\beta,
 \nonumber \\
\gamma_{\bar Q} 
&=&
\frac{7}{12}\gamma.\nonumber
\end{eqnarray}

{\it (a) Spiral-to-Paramagnet Transition.}
In principle, there are two ways in which the system can make a transition from the paramagnet into a spiral state:
\\ \\
i. {\it A second order transition in} $M$, along which $M=0$. This line is given by $\alpha_{\bar Q} =0$, and the necessary condition for its existence is that $\beta_{\bar Q} >0$. 
\\ \\
ii. {\it A first order transition in} $M$, along which $M$ jumps from zero to a finite value. This transition happens along the line $\beta_{\bar Q} ^2=4 \alpha_{\bar Q} \gamma_{\bar Q} $, as long as $\beta_{\bar Q} <0$ and $\alpha_{\bar Q} >0$. 
\\ 

\noindent
Since we have already established that we can have a spiral state only for $\beta_1<0$ and hence $\beta, \beta_{\bar Q} <0$ (following from the proportionalities of coefficients), we rule out the first possibility and conclude that the transition from the paramagnet into the spiral ferromagnet must be first order in $M$ (and also first order in $Q$, according to 
Eq.~(\ref{QM})). Substituting $\alpha_{\bar Q} $, $\beta_{\bar Q}$ and $\gamma_{\bar Q} $ from Eq.~(\ref{phiMspiral}), the equation for this line becomes $\alpha \gamma=\frac{17}{63}\beta^2$. This transition pre-empts the transition from the paramagnet into the uniform ferromagnetic state [see Fig.~\ref{spiraldiagram}].

{\it (b) Uniform Ferromagnet-to-Spiral Transition.}
Next, we wish to determine the boundary between the spiral phase and the uniform ferromagnet.
In principle, this may occur either discontinuously or smoothly. In the case of the free-electron dispersion, it occurs {\it via} a Lifshitz transition where $\bar Q$ goes continuously to zero. The value of magnetization that minimizes the free energy $\mathcal{F}_{\bar Q}$ is given from Eq.~(\ref{phiMspiral}) by
\begin{equation}
M^2=\frac{-2\beta}{7\gamma}\left (\frac{8}{9}+\sqrt{\left (\frac{8}{9}\right )^2-7\left ( \frac{\alpha \gamma}{\beta^2}-\frac{5}{27} \right) } \right ).
\end{equation}
Substituting this into Eq.~(\ref{QM}) for $\bar Q$, we find that the Lifshitz line coincides with line $\alpha=0$. The magnetization $M$ varies continuously along this line.

\begin{figure}[ht]
\includegraphics[width=0.75\linewidth]{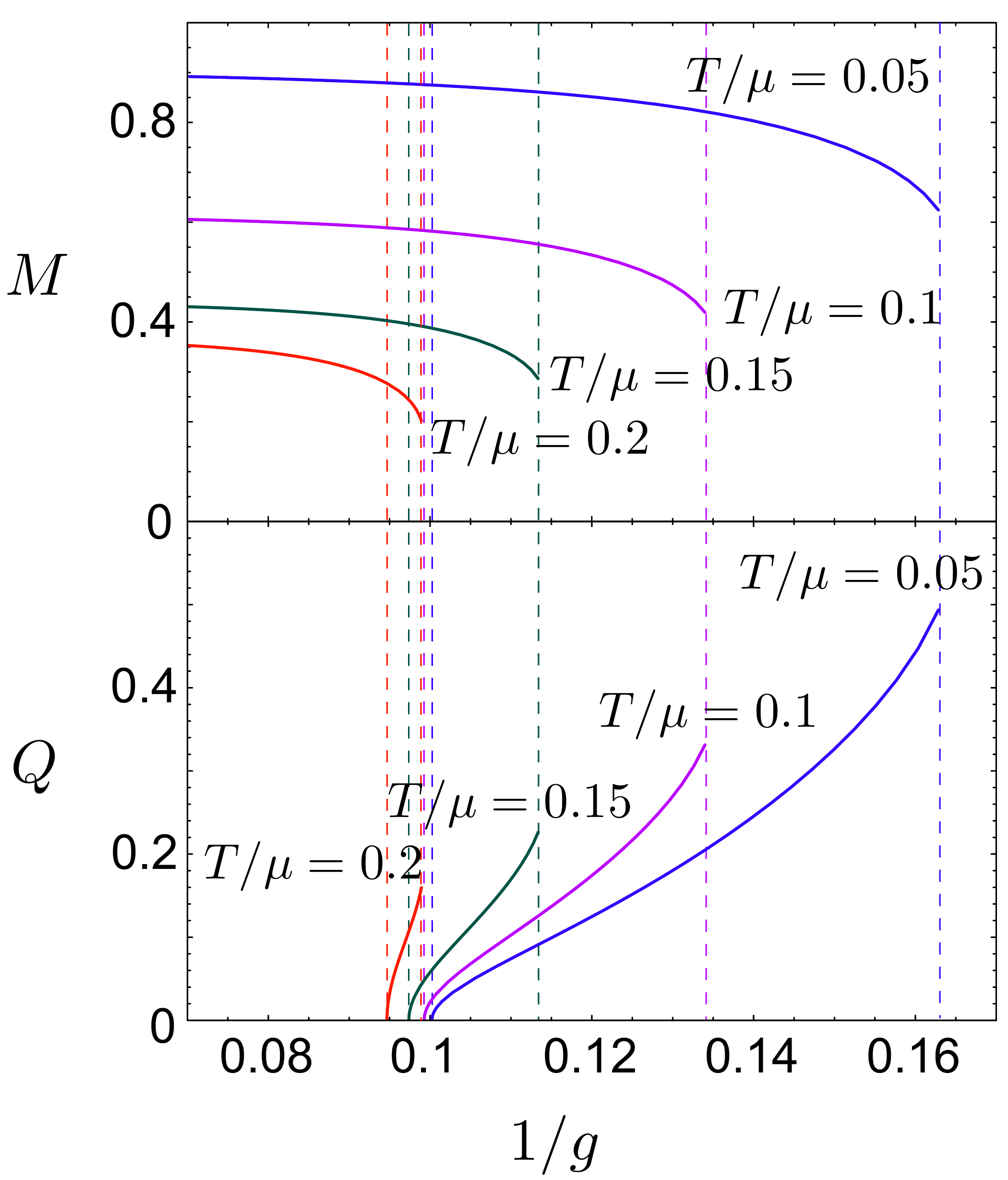}
\caption{(Color online) The evolution of the order parameters $M$ and $Q$ in the spiral phase. As we approach the tricritical point, the jumps in $M$ and $Q$ along the line of first order transitions become smaller. At the tricritical point $M=Q=0$. At the Lifshitz transition between the uniform ferromagnet and the spiral ferromagnet, $Q$ goes continuously 
to zero while $M$ remains finite and behaves smoothly.}
\label{mqevolution}
\end{figure}
The evolution of the order parameters $M$ and $Q$ in the vicinity of the first order transition from the paramagnet into the spiral state are plotted in Fig.~\ref{mqevolution}. We see that the jumps in $M$ and $Q$ get smaller as we approach the tricritical point. 

Previous analyses \cite{Conduit+09} determined the phase diagram of spiral and uniform ferromagnets (they did not consider a spin nematic phase) by brute force numerical (Monte-Carlo) evaluation of the fluctuation corrections given by Eq. (\ref{freeenergyn}) for  given $g$, $T$ and $M$,$Q$. We, instead, evaluate the phase diagram within a Ginzburg-Landau expansion and obtain an analytical approximation at low temperatures. The agreement between the two methods is good in the vicinity of the tricritical point where the expansion
is controlled. 
In addition, we were also able to determine the location of the boundary between the uniform and spiral ferromagnets as well as the nature of this transition.

\subsubsection{Fluctuation-Driven Spin Nematic}
Finally, we determine the region of the phase diagram where the d-wave spin-nematic phase has the lowest free energy.  By analyzing the free energy Eq.~(\ref{nematicfree}), we find that for temperatures below $T=0.02 \mu$, the transition from the paramagnet into the spiral state is pre-empted by a transition into a spin-nematic state. The first order transition line between the paramagnet and the spin nematic is given by the equation 
\begin{equation}
\beta^2
=
4 
\frac{\langle d_{{\bf k}}^2 \rangle\langle d_{{\bf k}}^6 \rangle}{\langle d_{{\bf k}}^4 \rangle ^2}
 (g-\alpha) \gamma. 
 \end{equation}
From the evaluation of this equation for spin-nematic states with d- and p-wave symmetry we also find that the instability to the formation of the d-wave spin nematic occurs 
at slightly higher temperatures and is therefore favored. This however might change with dimensionality, the form of the electron dispersion, or the range of the interactions. 

\begin{figure}[ht]
\includegraphics[width=0.95\linewidth]{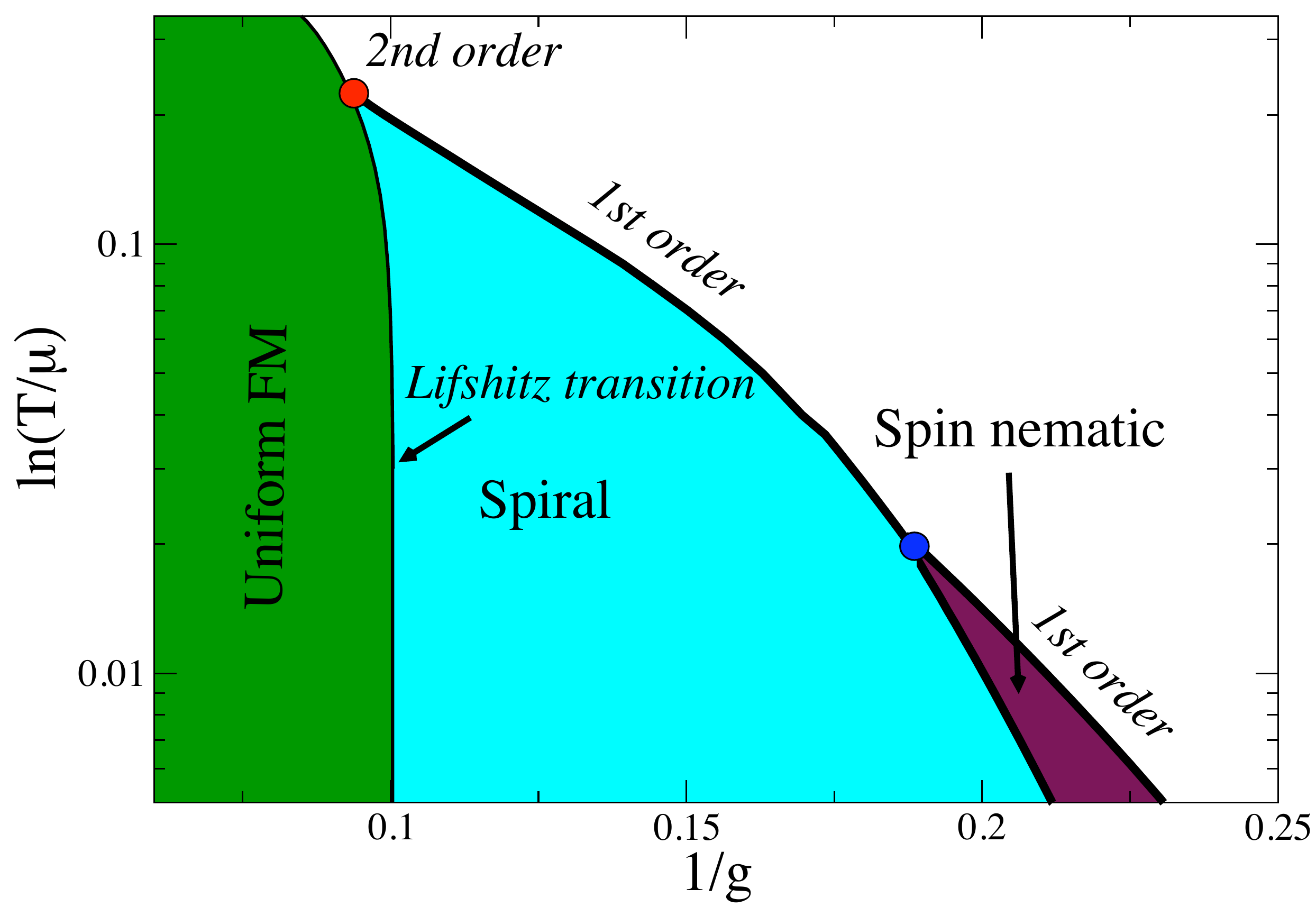}
\caption{(Color online) Phase diagram of the itinerant ferromagnet, allowing for the possibility for the formation of spiral and spin-nematic phases. At temperatures which are about
an order of magnitude smaller than the temperature of the tricritical point, a d-wave spin nematic forms between the spiral ferromagnet and the paramagnet.}
\label{nematicdiagram}
\end{figure}
Comparing the free energies of the spiral and spin-nematic phases, we find that the spin-nematic state penetrates into the region where the spiral state was previously favored. 
The details of the transition between the spiral and spin-nematic phases are potentially very interesting but hard to analyze. Introducing phase slips into the spiral restores translational symmetry and renders the phase nematic. Whether this is indeed the spin-nematic phase identified here, or something more exotic\cite{Senthil} is not clear. 

The phase diagram, including the spin-nematic state is shown in Fig.~\ref{nematicdiagram}. Note that this phase diagram is plotted on a logarithmic scale, since the spin-nematic state onsets at a temperature which is an order of magnitude lower than the temperature of the tricritical point where the spiral forms. 

In summary, quantum fluctuations have generated a coupling in the spin-nematic channel and stabilized a spin-nematic phase. This is similar to the mechanism by which a superconducting state is stabilized in spin-fluctuation theory,\cite{Anderson+73,Fay+80} and we anticipate that our approach can be applied to study superconductivity as well.
We emphasize that the quantum order-by-disorder approach incorporates charge fluctuations on the same footing as spin fluctuations (see Section II C).
As pointed out by Chubukov and Maslov,\cite{Chubukov+09} charge fluctuations are essential to mediate the formation of a spin-nematic state.

\subsection{Phase Diagram for Anisotropic Dispersion}
Changing from a free-electron dispersion to a band dispersion can have a profound effect upon the magnetic phase diagram. In the extreme, it can lead to nesting and the formation of modulated (anti-ferromagnetic) phases even at mean field. Here, we consider the effect of a weakly anisotropic dispersion
$\epsilon_{\bk}=\frac{k^2}{2}+\delta \epsilon_{\bk}$,
with $\delta \epsilon_{\bk}=\delta \cos{(4 \phi)} \sin{\theta}$. This deformation makes the dispersion more tight-binding like. By changing the subtle balance between kinetic  energy and fluctuation corrections the topology of the phase diagram, the regions occupied by the different phases, and the nature ({\it e.g.} first or second order) of the transitions are altered. 

\subsubsection{Uniform Ferromagnet}
First,  we investigate the changes to the phase diagram of the uniform ferromagnet. Summing Eqs. (\ref{hfmGL}) and (\ref{freehommods}) evaluated at $Q=0$, we arrive at the following expression for the free energy:
\begin{eqnarray}
\label{anisuniform}
\tilde  {\cal F}
&=&
\tilde \alpha M^2+\tilde \beta M^4 + \tilde\gamma M^6, 
 \\
\tilde \alpha 
&=&
\alpha+ \left ( 6 \beta_{\textrm{MF}}+\frac{g^2}{2}\frac{\partial^2 \alpha_{\textrm{fl}} }{\partial^2 \mu^2} \right ) \frac{\langle \delta \epsilon_{\bk} ^2\rangle }{g^2},
 \nonumber \\
\tilde \beta 
&=&
 \beta +15 \gamma \frac{\langle \delta \epsilon_{\bk} ^2\rangle}{g^2},\nonumber\\
 \tilde\gamma
 &=& 
 \gamma.\nonumber
\end{eqnarray}
As in the case of the isotropic $k^2$ dispersion, we find that the transition between the uniform ferromagnet and the paramagnet is continuous  at high temperatures
and becomes first order at low temperatures due to a sign change of $\tilde\beta$. The line of second order transitions between the uniform ferromagnet and the paramagnet 
is given by $\tilde \alpha =0$ while the line of first order transitions is given by $\tilde \beta ^2=4 \tilde \alpha \tilde\gamma$. The effect of the anisotropic correction to 
the dispersion is to slightly shift the locations of the phase boundaries, e.g. the temperature of the tricritical point is reduced to $T^*=0.225\mu$ (see Fig.~\ref{anisotropicphase}).

As we will see in the following, the effects of the anisotropy on the formation of the spiral and the nature of the transitions to the spiral ferromagnet are more interesting. 

\subsubsection{Fluctuation-Driven Spiral}
Our analysis of the spiral phase follows the same steps as in the case of the free-electron dispersion in Section IV.2. The resulting expressions are lengthy and not particularly revealing in themselves. Therefore, we simply outline the main steps. 
The free energy of the spiral state is the sum of Eqs.~(\ref{spiralfree}) and (\ref{freehommods}), and is given by
\begin{eqnarray}
\label{spiralbandprelim}
\mathcal{F}[M,{\bf Q}] & = &  
\left(\tilde\alpha+\tilde\beta_{1}(\hat{{\bf Q}}) Q^2+\tilde\gamma_{1} Q^4\right) M^2 \nonumber\\
& & +\left(\tilde\beta+\tilde\gamma_{2} Q^2 \right) M^4 
+\tilde\gamma M^6,
\end{eqnarray}
with $\tilde\alpha$, $\tilde\beta$, and $\tilde\gamma$ defined in Eq.~(\ref{anisuniform}), $\tilde\gamma_1=\gamma_1=\frac 35 \gamma$, $\tilde\gamma_2=\gamma_2= \gamma$, 
and
\begin{eqnarray}
\tilde \beta_1(\hat{{\bf Q}})
&=&
\frac{2}{3} \beta
+
\frac{\beta_{MF}}{2}
 \bigg\langle
  \left(\frac{\bQ\cdot\nabla \delta \epsilon_{\bk}}{k_F Q}\right)^2 
 \bigg\rangle 
 \nonumber\\
 & &
+ 30\frac{ \gamma}{g^2} 
\bigg\langle 
\left(\frac{\bk\cdot \bQ}{k_F Q}\right)^2 
\delta \epsilon_{\bk}^2
\bigg\rangle.
\end{eqnarray}
This free energy now depends upon the direction $\hat{{\bf Q}}={\bf Q}/Q$ and is no longer invariant under rotations of the spiral. This is the consequence 
of the anisotropic dispersion which breaks the continuous rotation symmetry. It turns out that for the particular anisotropy $\delta \epsilon_{\bk}=\delta \cos{(4 \phi)} \sin{\theta}$, the free energy is minimized for spirals with $\hat\bQ$ along the z-axis. 

\begin{figure}[ht]
\includegraphics[width=0.95\linewidth]{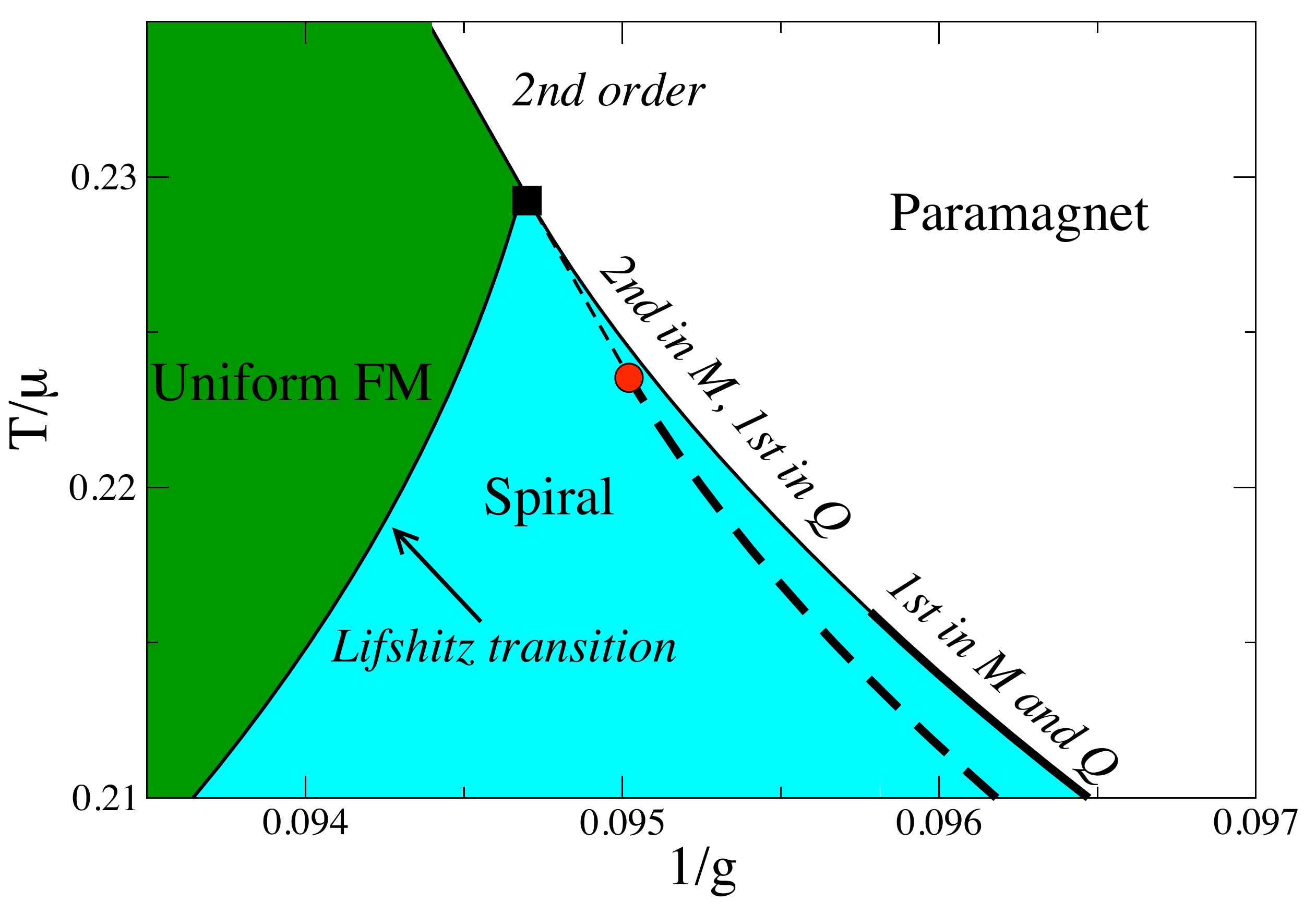}
\caption{(Color online) Modification to the phase diagram of the spiral state with a weakly anisotropic dispersion. The onset of the spiral no longer coincides with the 
uniform tricrital point (circle). Instead, the spiral forms at a slightly higher temperature (square) and pre-empts a portion of the continuous transition between the uniform ferromagnet 
and the paramagnet (thin dashed line) as well as the first-order transition (thick dashed line). Note that because of the anisotropy also the nature of the spiral-to-paramagnet 
transition changes. At higher temperatures $M$ now behaves continuously while at low temperatures the transition is first order in $M$ as in the isotropic case.}
\label{anisotropicphase}
\end{figure}

Notice that the proportionality between coefficients found in the case of the free-electron dispersion is broken by the anisotropic dispersion. For example, the coefficient of the term $Q^2 M^2$ is no longer proportional to that of the $M^4$ term. This broken symmetry changes the topology of  the phase diagram and changes the nature of the transition between the spiral ferromagnet and the paramagnet (see Fig.~\ref{anisotropicphase}). 

{\it (a) Optimum wavevector.}
The optimum wavevector is obtained by minimizing the free energy Eq.~(\ref{spiralbandprelim}) with respect to $Q$ for a given magnetization $M$. In this way, 
we obtain ${\bar Q} \equiv {\bar Q}[M]$. Substituting this value of $Q$ back into Eq.~(\ref{spiralbandprelim}) we obtain the free energy at the optimum wavevector as a function of $M$,
\begin{eqnarray}
\label{bandMspiral}
\tilde {\cal F}_{\bar Q}[M]
&=&
\tilde \alpha_{\bar Q} M^2+\tilde \beta_{\bar Q} M^4 + \tilde \gamma_{\bar Q} M^6 ,
\end{eqnarray}
where $\tilde \alpha_{\bar Q}$, $\tilde \beta_{\bar Q}$ and $\tilde \gamma_{\bar Q}$ are the appropriate generalizations of the free-electron forms given in Eq.~(\ref{phiMspiral}).

{\it (b) Paramagnet-to-Spiral Transition.}
As discussed in the case of the isotropic free-electron dispersion, in principle, there are two ways in which the system can make a transition form the paramagnet into the spiral state:
i. a second order transition in $M$ (which occurs along the line $\tilde \alpha_{\bar Q}=0$ for $\tilde \beta_{\bar Q} >0$), and
ii. a first order transition in $M$ (which occurs along the line 
$\tilde \beta_{\bar Q}^2 = 4 \tilde \alpha_{\bar Q} \tilde \gamma_{\bar Q}$ for $\tilde \beta_{\bar Q} <0$ and $\tilde \alpha_{\bar Q} >0$). 
For the free-electron dispersion only the latter possibility occurred. Now that we have broken the proportionality between Ginzburg-Landau coefficients by allowing for an anisotropic dispersion, both of the possibilities exist. 
\\
\noindent
i. The transition begins as second order in $M$ (and first order in $Q$) (thin solid line in Fig.~\ref{anisotropicphase}, extending below the square symbol). This line was not present in the phase diagram 
with free-electron dispersion, because 
the quartic coefficient $\beta$ was negative in the region where the spiral existed. 
This line pre-empts a portion of the line of second order phase transitions between the paramagnet and the uniform ferromagnet (thin dashed line) and the line of first order 
transitions from the paramagnetic to the uniform ferromagnetic state (thick dashed line). The formation of the spiral phase is, therefore, no longer coincident with the putative tricritical point of the uniform ferromagnet (shown as circle in Fig.~\ref{anisotropicphase}) but sets in at higher temperatures as indicated by a square symbol in Fig.~\ref{anisotropicphase}. 
\\
\noindent
ii. The second order transition between the paramagnetic and spiral phases gives way to a first order transition at lower temperatures - shown as thick solid line in Fig. \ref{anisotropicphase}.

{\it (c) Uniform Ferromagnet-to-Spiral Transition. }
The boundary between the spiral and uniform ferromagnetic phases remains a Lifshitz transition, where the optimal wave vector falls continuously to zero and $M$ is continuous. In the case of the free-electron dispersion, this has turned out to be coincident with the $\alpha=0$ line. The situation is not so simple when we allow for an anisotropic dispersion. 
While the magnetization remains continuous,  within our numerical resolution, we cannot exclude that the derivative of $M$ becomes discontinuous. 
The anisotropic dispersion has had two key effects upon the phase diagram. Firstly, the region occupied by the spiral phase has increased, and secondly, the onset of the spiral decoupled from the tricritical point of the uniform ferromagnet.
\subsubsection{Fluctuation-Driven Spin Nematic} 
The free energy of a d-wave spin-nematic state in the presence of the distortion is given by the sum of Eqs.~(\ref{nematicfree}) and (\ref{freehommodn}). We can rewrite this in the form
\begin{equation}
\tilde {\cal F}[N]=\tilde \alpha_N N^2 +\tilde \beta_N N^4 +\tilde \gamma_N N^6,
\end{equation}
where
\begin{eqnarray}
\tilde \alpha_N
&=&
- \langle d_{{\bf k}}^2 \rangle(\alpha-g) 
 - \left ( 
 6 \frac{\beta_{\textrm{MF}}}{g^2}+\frac{1}{2}\frac{\partial^2 \alpha_{\textrm{fl}} }{\partial \mu^2} 
 \right )
 \langle \delta \epsilon_{\bk} ^2 d_{\bk}^2 \rangle,
 \nonumber\\
 \tilde \beta_N
 &=&
\langle d_{{\bf k}}^4 \rangle\beta
+
15 \frac{\gamma}{g^2}
 \langle \delta \epsilon_{\bk} ^2d_{\bk}^4 \rangle,
\nonumber\\
\tilde \gamma_N
&=&
\langle d_{{\bf k}}^6 \rangle \gamma.
\end{eqnarray}
The transition from the paramagnet into the d-wave spin nematic is first order and occurs when 
$\tilde  \beta_N ^2=4 \tilde \alpha_N \tilde \gamma_N$.
The boundary between spiral and nematic states is obtained by comparison of their free energies - although, as we stated above, the details of how this transition occurs may be subtle. When the dispersion is anisotropic, the spin nematic state occurs at higher temperatures than in the isotropic case. The spin nematic is 
stabilized since it becomes easier to redistribute the kinetic energy cost of forming a spin nematic when the dispersion is anisotropic.

In summary, we have used the Ginzburg-Landau expansion of the free energy, including the quantum fluctuations, to determine the phase diagram of the itinerant ferromagnet in the vicinity of a quantum critical point. First, we investigated the effect of quantum fluctuations upon the phase diagram of the uniform ferromagnet. Below a certain temperature (the tricritical temperature) the paramagnet-to-ferromagnet transition becomes first order. Next, we allowed for the possibility of a spatially modulated ferromagnetic phase. For temperatures lower than the tricritical temperature, it becomes energetically favorable to form a spiral state in between the paramagnetic and uniform ferromagnetic states. 
The putative first order transition between the paramagnet and the uniform ferromagnet is pre-empted by a transition into this spatially modulated state. Further, we included the possibility of forming a d-wave spin-nematic state. This state is stabilized for temperatures, below $T\approx0.1 T^*\approx 0.02\mu$, in between the paramagnetic and spiral phases. Finally, we have shown that in the presence of an anisotropic dispersion, the topology of the phase diagram changes such that both spiral and spin-nematic phases occupy a larger region of the phase diagram. Moreover, the onset of spiral order no longer coincides with the uniform tricritical point.

\section{Conclusions and Outlook}
We have shown how quantum fluctuations can lead to the formation of new phases in the vicinity of itinerant ferromagnetic quantum critical points. Quantum order-by-disorder not only provides an intuitive physical picture for this process but identifies a general principle\cite{Laughlin+01} behind the phase reconstruction near quantum-critical points.

Quantum order-by-disorder is familiar in condensed-matter\cite{Mila+91,Zaanen00,Zaanen+01,Kruger+06} as well as in high-energy physics where it is referred 
to as the Coleman-Weinberg mechanism.\cite{Coleman+73} In these familiar realizations, new ground states are stabilized by quantum fluctuations of a bosonic order parameter.
What makes our approach new is the immediate connection between Fermi-surface deformations (associated with the onset of competing order) and the enhancement of
phase space available for low-energy quantum fluctuations. 

Recently, it has been argued\cite{She+10} that the avoidance of naked quantum-critical points due to the quantum order-by-disorder mechanism  can be understood within the 
AdS/CFT correspondence\cite{Maldacena98,Witten98,Gubser+98} between conformal field theories, describing critical condensed-matter systems, and gravity in Anti-de-Sitter space. 
In the gravity context, the quantum-critical state at finite temperatures corresponds with a Reissner-Nordstrom black hole in AdS space. It has been realized that such a black hole 
can become unstable at low temperatures and tends to collapse to a state with lower entropy.\cite{Hartnoll+08} This entropic motive underlies the quantum order-by-disorder mechanism. Experimentally, the measurement of entropic landscapes has proven a revealing probe of the phase reconstruction near quantum critical points.\cite{Borzi+07,Rost+09}

The fermonic quantum order-by-disorder approach presented in this paper not only establishes the connection to deformations of the Fermi surface, which are accessible by 
various experimental probes, but also leads to relatively simple analytical calculations, based on self-consistent second order perturbation theory. As such, it is more accessible than technically involved diagrammatic techniques.\cite{Belitz+97,Betouras+05,Rech+06,Efremov+08,Maslov+09} The two approaches are formally equivalent;  expanding self-consistently about a saddle point with the already established order re-sums selected series of diagrams that give rise to non-analytic corrections to the free energy.

We have used this approach to investigate the fluctuation-driven phase reconstruction in the vicinity of the itinerant ferromagnetic quantum critical point in three spatial 
dimensions. This quantum-critical point is unstable towards the formation of spiral and spin-nematic states. Quantum fluctuations would render the transition between
the uniform ferromagnet and the paramagnet first order. This first-order transition is pre-empted by a modulated, spiral ferromagnetic phase. At even lower temperatures, 
a d-wave spin-nematic state forms which is slightly favored over a spin-nematic with p-wave symmetry. It is sandwiched between the paramagnetic and spiral phases. In order to describe more generic experimental systems, we determined the topology of the phase diagram in the presence of an anisotropic electron dispersion. The regions of phase space occupied by both the spiral and spin nematic phases are enlarged. Moreover, the onset of spiral order and the putative tricritical point of the uniform ferromagnet become decoupled and the order of transitions is modified.

Similar spin-nematic instabilities near to the itinerant ferromagnetic quantum critical point have been studied recently by Chubukov and Maslov\cite{Chubukov+09} within 
an extension to Hertz-Millis theory in a spin-fermion model. The authors point out that the inclusion of the effects of charge fluctuations (through Aslamov-Larkin corrections) 
in addition to spin fluctuations is crucial to mediate the formation of a spin nematic. Our quantum order-by-disorder approach incorporates charge fluctuations on 
the same footing as spin fluctuations. The results of Ref.~[\onlinecite{Chubukov+09}] are very much in accord with those that we present here; a spin nematic instability 
occurring out of the paramagnetic state is found to pre-empt the spiral phase. We point that the contact interaction used here as opposed to the finite-range interaction in Ref.~[\onlinecite{Chubukov+09}] lowers the free energy of the spiral relative to the spin nematic.

The quantum order-by-disorder approach can be applied to a variety of systems and phases. Adding a small spin-orbit coupling to the Stoner model of magnetism 
had previously enabled us to explain the partially ordered phase of MnSi\cite{Pfleiderer+04} in terms of quantum order-by-disorder.\cite{Kruger+12} Work on the superconducting instability is in progress. It appears that our approach recovers the results of spin-fluctuation theory, thus revealing the link between spin-fluctuation theory used to treat superconducting 
instabilities, and the extensions of Hertz-Millis theory that explain the instability of quantum critical points to other types of order. 
 
 There are several natural directions for developing our approach. The inclusion of superconducting instabilities is a priority - the nature of the superconducting phase where it overlaps with the spatially modulated magnetic phases raises the intriguing possibility of spontaneous, fluctuation-driven, modulated superconductivity. The calculations 
 described in this paper are in three dimensions. Extending them to two dimensions, where the non-analytic effects of fluctuations are even stronger, is an important step. 
 Finally, the role of fluctuations near to anti-ferromagnetic quantum critical points may also be susceptible to analysis using our methods.

 \textbf{Acknowledgment:}  The authors benefited from stimulating discussions with 
 D. Belitz, A.~V. Chubukov,  G.~J. Conduit,  B.~D. Simons,  and J. Zaanen. 
 This work was supported by EPSRC under grant code EP/I 004831/1.

\appendix
\section{Free energy of the uniform ferromagnet in terms of modified particle-hole densities of states}\label{appendix1}

The fluctuation corrections to the free energy are given by a high dimensional integral over 
momenta $\bk_1,\ldots,\bk_4$ and correspond to excitations of virtual pairs of particle-hole pairs of opposite spin and equal and opposite momenta.  
It is therefore possible to rewrite the regularized fluctuation corrections $\mathcal{F}_\textrm{fl}$ (\ref{freeenergyn}) as a lower dimensional integral over modified particle-hole densities of states, 
\begin{equation}
\label{freeenergymodifieddos}
\mathcal{F}_\textrm{fl} =2 g^2\sum_{\sigma=\pm 1} \int_{\bq,\epsilon_1,\epsilon_2} \frac{\Delta\rho^\sigma(\bq,\epsilon_1)\rho^{-\sigma}(-\bq,\epsilon_2)}{\epsilon_1+\epsilon_2},
\end{equation}
where we have defined $\int_\bq:=\int \frac{ \ud^3 \bq}{(2 \pi)^3}$ and $\int_\epsilon:=\int_{-\infty}^\infty \ud\epsilon$. The modified particle-hole densities of states as a function of momentum $\bq$ and energy $\epsilon$ are given by

\begin{eqnarray}
\rho^\sigma(q,\epsilon) & = & \int_\bk n(\epsilon_{\bk-\frac{\bq}{2}}^\sigma)\delta(\epsilon-\epsilon_{\bk+\frac{\bq}{2}}^\sigma+\epsilon_{\bk-\frac{\bq}{2}}^\sigma),\\
\Delta \rho^\sigma(q,\epsilon) & = & \int_\bk n(\epsilon_{\bk-\frac{\bq}{2}}^\sigma)n(\epsilon_{\bk+\frac{\bq}{2}}^\sigma)\delta(\epsilon-\epsilon_{\bk+\frac{\bq}{2}}^\sigma+\epsilon_{\bk-\frac{\bq}{2}}^\sigma), \nonumber
\label{moddos}
\end{eqnarray}
and are related to the particle-hole density of states as $\rho^\textrm{ph}_\sigma=\Delta\rho_\sigma-\rho_\sigma$. This form of the fluctuation correction will prove useful in our subsequent evaluation of the Ginzburg-Landau expansion of the uniform ferromagnet, since the modified particle-hole density of states of the uniform ferromagnet can be calculated analytically. This leads to a tremendous simplification of the fluctuation integral.  

The modified particle-hole densities of states are functions of the magnetization $M$, which enters through the dispersion $\epsilon_\bk^\sigma=\epsilon_\bk-\sigma g M$ of the uniform ferromagnet. We want to Taylor expand  Eq. (\ref{freeenergymodifieddos}) with respect to $M$. In doing so we will require the derivatives of $\rho^\sigma$ and 
$\Delta \rho^\sigma$ with respect to $M$. However, since in $\rho^\sigma$ and $\Delta \rho^\sigma$ the dispersion only enters for either spin up or spin down (and not both) we can relate the derivatives with respect to $M$ to derivatives with respect to the chemical potential $\mu$,
\begin{eqnarray}
\left.
\partial^i_M \Delta \rho^\sigma(q,\epsilon)\right|_{M=0}
& = & 
\left.
(\sigma g)^i\partial^i_{\mu} \Delta \rho^\sigma(q,\epsilon)
\right|_{M=0}\nonumber\\
& = & 
(\sigma g)^i\partial^i_{\mu} \Delta \rho(q,\epsilon).
\end{eqnarray}
Now let us derive explicit expressions for $\Delta \rho=\left.\Delta \rho^\sigma\right|_{M=0}$ and $\rho=\left.\rho^\sigma\right|_{M=0}$ and their derivatives. 

\subsection{Evaluation of $\rho (q,\epsilon)$ and $\Delta \rho (q, \epsilon)$}
For an isotropic free-energy dispersion, $\epsilon_\bk=\frac 12 k^2$ the angular integrals in $\Delta \rho (q, \epsilon)$ and $\rho (q, \epsilon)$ are easy to compute in three dimensions since the only angular dependencies enter through the volume element and the scalar product $\bk\cdot\bq=kq\cos\theta$. The remaining one-dimensional integrals over $k$ are elementary.
The resulting modified particle-hole densities of states are given by

\begin{eqnarray}
\rho(q,\epsilon) & = & \frac{1}{(2\pi)^2}\frac{T}{q} \ln{\left( 1+e^{-\frac{1}{T}[\phi^+(\epsilon,q)-\mu]} \right)},\\
\Delta \rho(q,\epsilon)&=&\frac{1}{(2\pi)^2}\frac{T}{q}\left[ \frac{1}{1-e^{\frac{\epsilon}{T}}} \ln{\left( 1+e^{-\frac{1}{T}[\phi^-(\epsilon,q)-\mu]} \right)}\right.\nonumber \\
& &\;\;\; +\left.\frac{1}{1+e^{-\frac{\epsilon}{T}}} \ln{\left( 1-e^{-\frac{1}{T}[\phi^+(\epsilon,q)-\mu]} \right)}  \right],\nonumber
\end{eqnarray}
where,
\begin{eqnarray}
 \phi^{\pm}(\epsilon,q)=\frac{1}{2}\left(\frac{\epsilon}{q}\pm\frac{q}{2}   \right)^2.
\end{eqnarray}

For the derivatives of the modified densities of states with respect to the chemical potential we obtain

\begin{eqnarray}
\partial_\mu^i \rho(q,\epsilon) & = & \frac{1}{(2\pi)^2}\frac{1}{q}\partial_\mu^{(i-1)}n\left[ \phi^{+}(\epsilon,q)   \right],\\
\partial_\mu^i \Delta\rho(q,\epsilon)& = & \frac{1}{(2\pi)^2}\frac{1}{q}\partial_\mu^{(i-1)}\left\{ n\left[ \phi^{-}(\epsilon,q)   \right] \right. \left. n\left[  \phi^{+}(\epsilon,q)   \right]\right\},\nonumber
\end{eqnarray}
where $n(x)=1/[e^{(x-\mu)/T}+1]$ denotes the Fermi function.

\section{Fluctuation contributions to $\alpha$ and $\beta$}

%
\begin{figure}[ht]
\includegraphics[width=0.7\linewidth]{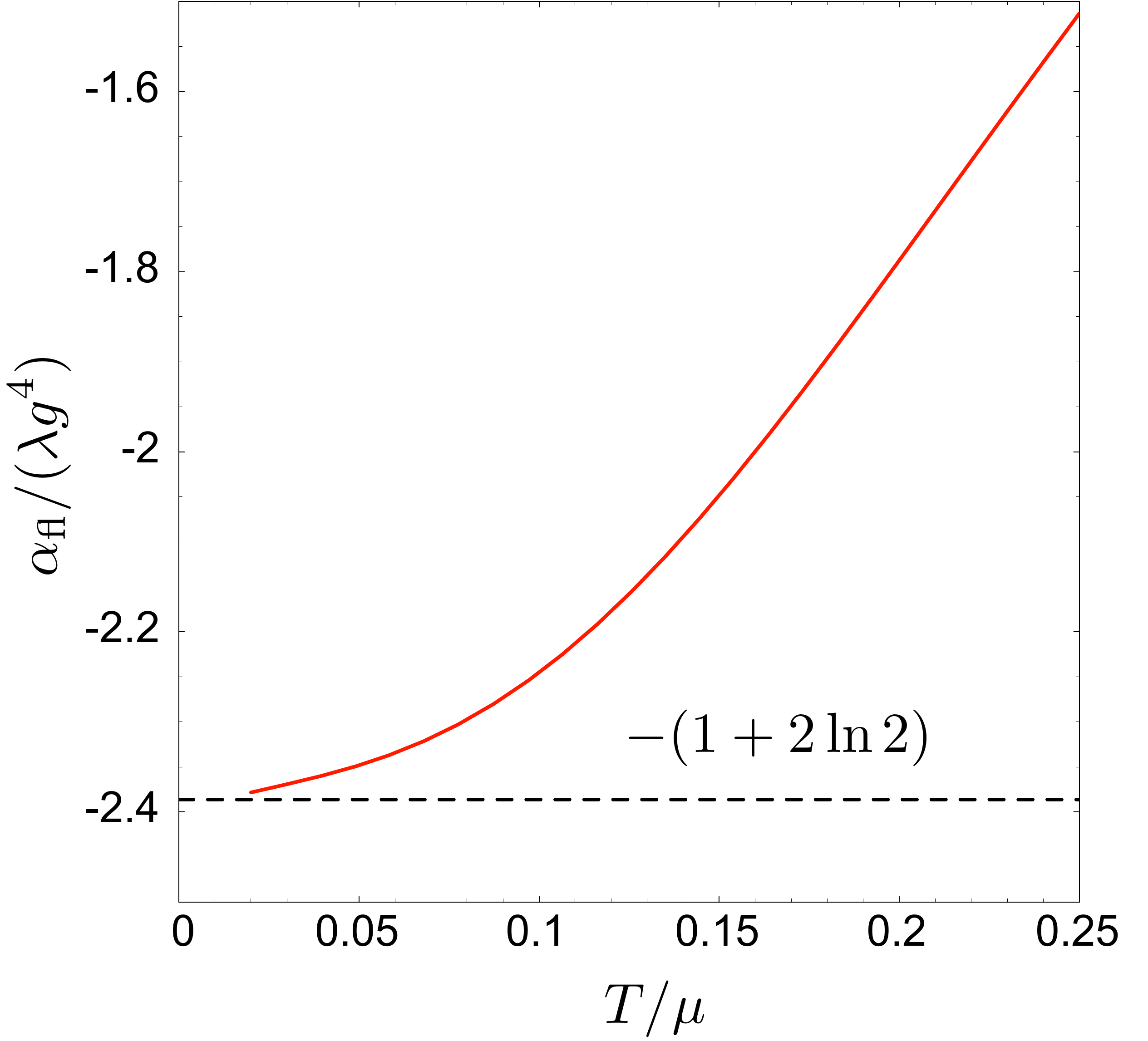}
\caption{(Color online) Fluctuation contributions to $\alpha$: comparison of numerics (solid line) with leading low-temperature analytical dependence (dashed line).}
\label{alpha}
\end{figure}

Here, we outline the evaluation of the fluctuation contributions to the Ginzburg-Landau coefficients $\alpha$ and $\beta$ of the uniform ferromagnet,
\begin{equation}
\alpha_{\textrm{fl}}=\frac{1}{2}\left.\frac{\partial^2 \mathcal{F}_{\textrm{fl}}}{\partial M^2}\right|_{M=0}\textrm{and}\quad
\beta_{\textrm{fl}}=\frac{1}{4!}\left.\frac{\partial^4 \mathcal{F}_{\textrm{fl}}}{\partial M^4} \right|_{M=0}.
\end{equation}
The modified particle-hole densities of states and their derivatives with respect to the chemical potential were calculated in Appendix \ref{appendix1}. For brevity, we define the 
integrals
\begin{equation}
J_{i,j} =  \int_{\bq,\epsilon_1,\epsilon_2}\frac{\partial_\mu^i \Delta\rho(q,\epsilon_1)\partial_\mu^j\rho(q,\epsilon_2)}{\epsilon_1+\epsilon_2}.
\end{equation}
The fluctuation contribution to the free energy at zero magnetization $\mathcal{F}_\textrm{fl}^{(0)}=\left.\mathcal{F}_\textrm{fl}\right|_{M=0}$ is given by
\begin{equation}
\mathcal{F}_\textrm{fl}^{(0)}=4 g^2 J_{0,0}.
\end{equation}
We evaluated the integral $J_{0,0}$ numerically for finite temperatures and analytically at $T=0$. The analytical calculation at $T=0$ correctly reproduces the result of 
Abrikosov and Khalatnikov.\cite{Abrikosov+58}
By Taylor expanding the fluctuation corrections to the free energy, we obtain the fluctuation contributions to the Ginzburg-Landau coefficients,
\begin{eqnarray}
\label{alphabetaJ}
\alpha_{\textrm{fl}} & = & 2g^4\sum_{i=0}^2 (-1)^i \left(\begin{array}{c} 2 \\ i   \end{array}\right) J_{i,2-i},  \nonumber\\ 
\beta_{\textrm{fl}} & = & \frac{g^6}{6}\sum_{i=0}^4 (-1)^i  \left(\begin{array}{c} 4 \\ i   \end{array}\right) J_{i,4-i}.
\end{eqnarray}
Some of the integrals $J_{i,j}$ are difficult to evaluate numerically, since at very low temperatures the higher derivatives of the Fermi functions (which enter through the derivatives of modified particle-hole densities of states) are strongly peaked around the Fermi energy and rapidly change sign.

\subsection{Evaluation of $\alpha$} 
From Eq. (\ref{alphabetaJ}) we see that we need to calculate three terms $J_{0,2},J_{2,0}$ and $J_{1,1}$. In principle, $J_{0,2},J_{2,0}$ are more difficult to calculate numerically (but possible) since they contain first derivatives of Fermi functions which are strongly peaked. However, we can reduce the number of terms that we need to calculate by collecting some together to re-express them as  derivatives of  $\mathcal{F}_\textrm{fl}^{(0)}$ with respect to $\mu$. For example,
\begin{equation}
\partial_\mu^2\mathcal{F}_\textrm{fl}^{(0)}=4g^2(J_{0,2}+2 J_{1,1}+J_{2,0}).
\end{equation}
%
Using these relations we can rewrite the fluctuation contribution to $\alpha$ as
\begin{eqnarray}
\label{alphaanalytics}
\alpha_{\textrm{fl}} 
&=&
2 g^4 \left( J_{0,2}-2J_{1,1}+J_{2,0}\right)
\nonumber\\
&=&
\frac{g^2}{2!} \partial_\mu^2 \mathcal{F}_\textrm{fl}^{(0)}-8 g^4J_{1,1}.
\end{eqnarray}
The remaining integral $J_{1,1}$ is easy to evaluate numerically. The temperature dependence of $\alpha_{\textrm{fl}}$ is shown in Fig.~\ref{alpha}. At zero temperature we obtain
\begin{equation}
\alpha_\textrm{fl} \simeq -\lambda (1+2\ln2) g^4,
\end{equation}
where $\lambda=\frac{16\sqrt{2}}{3 (2\pi)^6}$.

\subsection{Evaluation of  $\beta$}

\begin{figure}[ht]
\includegraphics[width=0.7\linewidth]{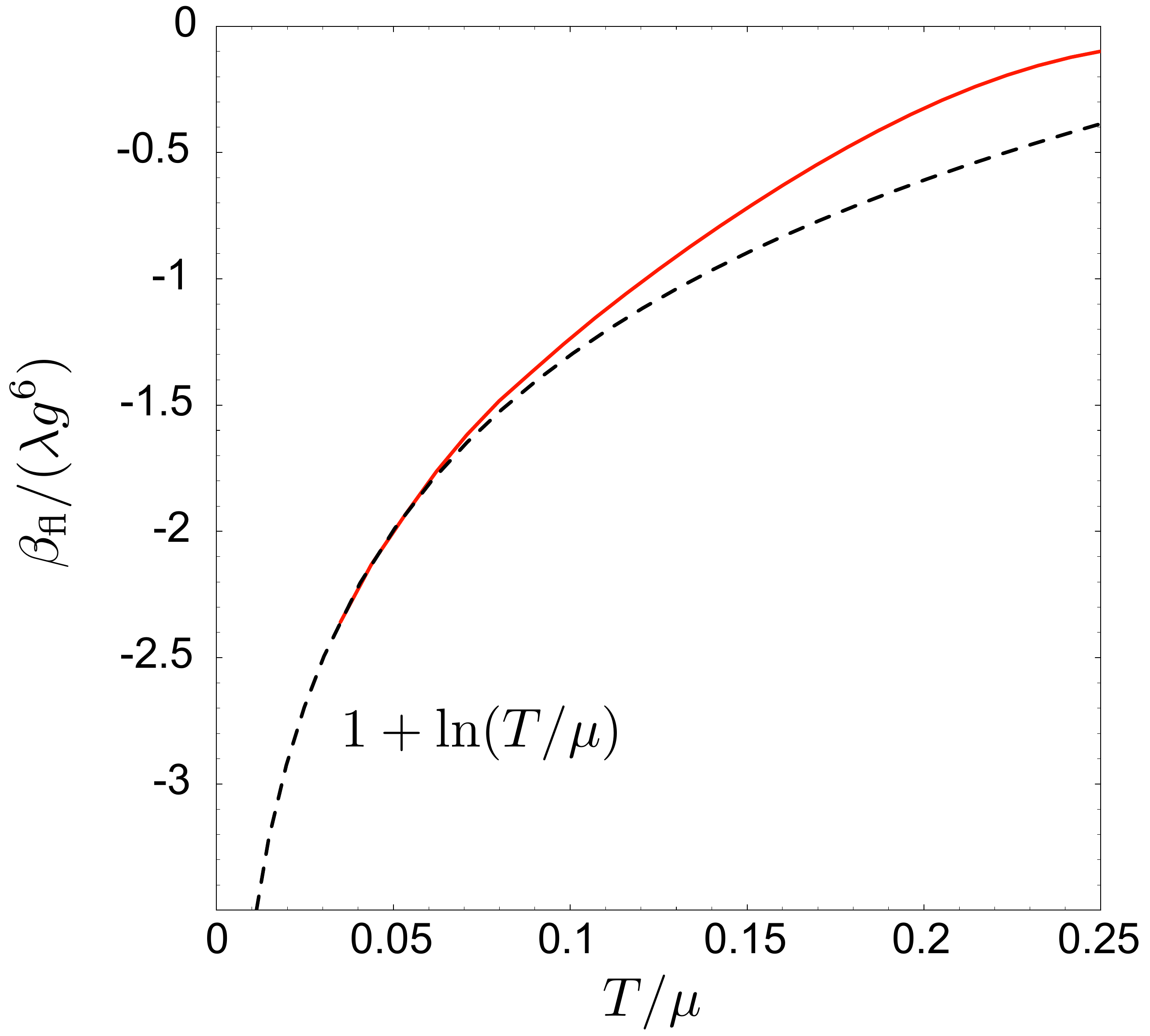}
\caption{(Color online) Fluctuation contributions to $\beta$: comparison of numerics (solid line) with leading low-temperature analytical dependence (dashed line).}
\label{beta}
\end{figure}

In order to evaluate $\beta_{\textrm{fl}}$ we need to evaluate five terms in Eq. (\ref{alphabetaJ}). As they contain higher order derivatives of Fermi functions 
they are even more difficult to evaluate numerically. We collect some of the terms together by noting that
\begin{eqnarray}
\partial_\mu^4\mathcal{F}_\textrm{fl}^{(0)}=4g^2\sum_{i=0}^4  \left(\begin{array}{c} 4 \\ i   \end{array}\right) J_{i,4-i},
\end{eqnarray}
to get
\begin{equation}
\label{betafJ}
 \beta_{\textrm{fl}}=\frac{g^4}{4!} \partial_\mu^4 \mathcal{F}_\textrm{fl}^{(0)}-\frac{4 g^6}{3}(J_{1,3}+J_{3,1}).
\end{equation}
In this way we avoid the explicit calculation of integrals $J_{0,4}$ and $J_{4,0}$, which contain third derivatives of Fermi functions. We can further simplify by noting that 
\begin{eqnarray}
\label{J13}
\partial_\mu^2 J_{1,1}=J_{3,1}+2 J_{2,2}+J_{1,3}.
\end{eqnarray}
If we re-express $J_{3,1}+J_{1,3}$ in Eq. (\ref{betafJ}) using Eq. (\ref{J13}), we obtain
\begin{equation}
 \beta_{\textrm{fl}}=\frac{g^4}{4!} \partial_\mu^4 \mathcal{F}_\textrm{fl}^{(0)}-\frac 43 g^6  \partial_\mu^2 J_{1,1}+\frac 83  g^6J_{2,2}.
\end{equation}
We have already calculated the functions $\mathcal{F}_\textrm{fl}^{(0)}$ and $J_{1,1}$, when we evaluated $\alpha_{\textrm{fl}}$. Both are smooth functions and the 
numerical evaluation of the derivatives with respect to $\mu$ is trivial. The leading temperature dependence of 
$\beta_{\textrm{fl}}$ comes from the $J_{2,2}$ integral, which diverges as $T\rightarrow 0$. In this limit
\begin{eqnarray}
\beta_\textrm{fl} &\simeq & \lambda \left (1+\ln{\frac T\mu} \right ) g^6.
\end{eqnarray}
The logarithmic divergence with temperature  arises from  particle-hole pairs with momenta $q \approx 2 k_F$. The good agreement between our numerical and analytical results in the low-temperature regime is shown in Fig.~\ref{beta}.

\end{document}